\RequirePackage{fix-cm}
\documentclass[smallextended]{svjour3}       
\smartqed  
\usepackage{graphicx}
\graphicspath{{}{../}}
\usepackage{color}
\usepackage{amsmath}
\usepackage{float}
\usepackage{array}
\usepackage{todonotes}

\begin{document}

\title{Are homeostatic states stable?\\ Dynamical stability in morphoelasticity }

\author{Alexander Erlich \and Derek E. Moulton \and Alain Goriely}


\institute{A. Erlich \at
              School of Mathematics, University of Manchester, Oxford Road, Manchester M13 9PL
           \and
           D. E. Moulton  \and A. Goriely \at
              Mathematical Institute, University of Oxford, Andrew Wiles Building, Woodstock Road, Oxford OX2 6GG
}


\maketitle

\begin{abstract}
Biological growth is often driven by mechanical cues, such as changes in external pressure or tensile loading. Moreover, it is well known that many  living  tissues actively maintain a preferred level of mechanical internal stress, called the mechanical homeostasis. The tissue-level feedback mechanism by which changes of the local mechanical stresses affect growth is called a growth law within the theory of morphoelasticity, a theory for understanding the coupling between mechanics and geometry in growing and evolving biological materials. 
The goal of this article is to develop mathematical techniques to analyze growth laws and to explore issues of heterogeneity and growth stability.  We discuss the growth dynamics of tubular structures, which are very common in biology (e.g. arteries, plant stems, airways) and model the homeostasis-driven growth dynamics of tubes which produces spatially inhomogeneous residual stress. We show that the stability of the homeostatic state  depends nontrivially on the anisotropy of the growth response. The key role of anisotropy may provide a foundation for experimental testing of homeostasis-driven growth laws. 
\end{abstract}

\vspace{5mm}

Biological tissues exhibit a wide range of mechanical properties and active behavior. A striking example is biological growth in response to the tissues mechanical environment. Artery walls thicken in response to increased pressure \cite{be74,Goriely2010}, axons can be grown by applying tension \cite{lazhbu92,rejego16}, and plant growth is driven by various mechanical cues \cite{gorota07,bo10}. The general idea underlying these phenomena is that the internal stress state is a stimulus for growth. As stress is rarely uniform, mechanically induced growth often coincides with differential growth, in which mass increase occurs non-uniformly or in an anisotropic fashion. In turn, differential growth produces residual stress, an internal stress that remains when all external loads are removed, appearing due to geometric incompatibility induced by the differential growth. Residual stress has been observed in a number of physiological tissues, such as the brain \cite{Budday2014}, the developing embryo \cite{Beloussov2006}, arteries \cite{fuli89}, blood vessels \cite{fu91}, solid tumors \cite{mapl04}, and in a wealth of examples from the plant kingdom \cite{goriely17}.  In many cases, residual stress has been found to serve a clear mechanical function; for instance in regulating size and mechanical properties.

Many living tissues actively grow in order to maintain a preferred level of internal residual stress, termed mechanical homeostasis. This phenomenon is characterized by growth being induced by any difference between the current stress in the tissue and the preferred homeostatic stress. Mechanically driven growth towards homeostasis poses several interesting and important questions, at the biological, mechanical, and mathematical level.
For instance, what determines the homeostatic stress state? At the cellular level, the growth response may be genetically encoded, with a homeostatic state manifest by differential cellular response to mechanical stimuli. From a continuum mechanics point of view, a residually stressed configuration is typically thought of as corresponding to a deformation from an unstressed configuration; however, it is not clear that such a deformation should exist to define a homeostatic state. Connected to this is a question of compatibility: is it actually {\it possible} for a system to reach mechanical homeostasis? For example, the boundary of an unconstrained tissue will always be traction free, and thus if the homeostatic stress for those boundary cells is non-zero, then the system can never completely reach homeostasis. From a dynamics point of view, there is a natural question of stability: is the homeostatic state stable, i.e. if the system is perturbed from its homeostatic equilibrium, is it able to grow in such a way to return to this state? There is also a practical issue of connecting experiment to theory: how does one quantify the homeostatic state and form of growth response? 

Mathematical modeling can be of significant value in addressing such questions and in suggesting potential experimental measures to quantify the properties of homeostasis. In the simplest and most widely used form, the mathematical description involves a growth law of the form
\begin{equation}\label{eq1}
\mathbf{G}^{-1}\dot{\mathbf{G}} = \boldsymbol{K}:(\mathbf{T}-\mathbf{T^*}).
\end{equation}
Here overdot represents time derivative, $\mathbf{G}$ is a growth tensor, characterizing the increase or decrease in mass as a local property, $\mathbf{T}$ is the Cauchy stress tensor, $\mathbf{T^*}$ is the preferred homeostatic stress tensor, and $\boldsymbol{K}$ is a fourth order tensor characterizing the growth response rate due to differences in current and preferred stress. Laws of the form \eqref{eq1}, or slight variations thereof, in which growth is coupled to Cauchy stress have been examined by a number of authors \cite{vago09,Bowden:2015kn,Ramasubramanian2008,Taber2008}, though the most appropriate form of growth law is a much-debated issue \cite{Taber1995,Ambrosi2011,Jones2012,goriely17}. An alternative but related approach involves coupling growth and Eshelby stress \cite{Ambrosi2005} based on thermodynamical arguments \cite{Epstein2000,erlich2015short}. Attempts to restrict the form of growth laws through thermodynamical considerations such as the Coleman-Noll procedure \cite{Coleman1963} have been of limited success due to the inherent thermodynamical openness and non-equilibrium nature of biological systems \cite{Maugin1999,Lebon2008}. The integration of micro-mechanical models with tissue level modeling has also been difficult, partly because the lack of periodicity and crystal symmetry in biological tissues makes the application of homogenization techniques difficult \cite{Chenchiah2014}.  Growth dynamics that depend on the current stress state are inherently challenging to study analytically. Both stress and growth will tend to be spatially dependent, with stress being determined through the solution of a force balance boundary-value problem, and thus any model will by nature involve a partial differential equation system. The situation is simplified somewhat by the {\it slow-growth assumption}, which states that growth occurs on a much longer time scale than the elastic time scale and hence the system is always in a quasi-static mechanical equilibrium.

In this paper we study mechanically driven growth in the context of growing tubular structures. One motivation for a cylindrical geometry is that such structures are ubiquitous in the biological world, from plant stems \cite{gomova10}  to axons and airways \cite{mogo11,mogo11b}, and exhibit diverse mechanical behavior.
Working within a constrained geometry will also enable us to gain qualitative insight into the dynamics of structures with growth driven by mechanical homeostasis, and to formulate a basic framework for studying the stability of a homeostatic state.
Even in an idealized geometry, the full growth dynamics still consists of a set of partial differential equations, with mechanical equilibrium requiring the solution of a boundary-value problem at each time step, and a highly nonlinear growth evolution for components of the growth tensor. There is no mathematical theory, yet, that allows for such an analysis. Our approach is therefore to devise a discretization though a spatial averaging scheme that converts the system to a much more manageable initial-value problem, to which we can apply standard techniques from dynamical systems. The discretization we propose consists of defining annular layers of the tubular structure, such that growth is uniform in each layer, driven by averaged values of the stress components in a law of the form \eqref{eq1}. While this approach enables us to  study efficiently properties of the continuous (non-discretized) system as the number of layers increases, for a smaller number of layers it is also a useful model of a multi-layered tube commonly found in many biological systems.

This paper is structured as follows. In Section \ref{sec:General_disks} we discuss the general deformation and growth dynamics for a tubular structure that is  homogeneous in the axial direction. In Section \ref{sec:two_disks} we focus on a tubular system made of two layers, illustrating the main ideas of our discretization approach and illustrating the rich dynamics of this system. In Section \ref{sec:N_disks} we generalize from two to $N$ layers. Here we find a rapid convergence of behavior as the number of layers increases, and investigate how the anisotropy of the growth affects the stability.

\section{\label{sec:General_disks}Continuous growth dynamics in cylindrical
geometry}

\subsection{Kinematics}

\begin{figure}[htpb]
\centering\includegraphics[width=0.7\textwidth]{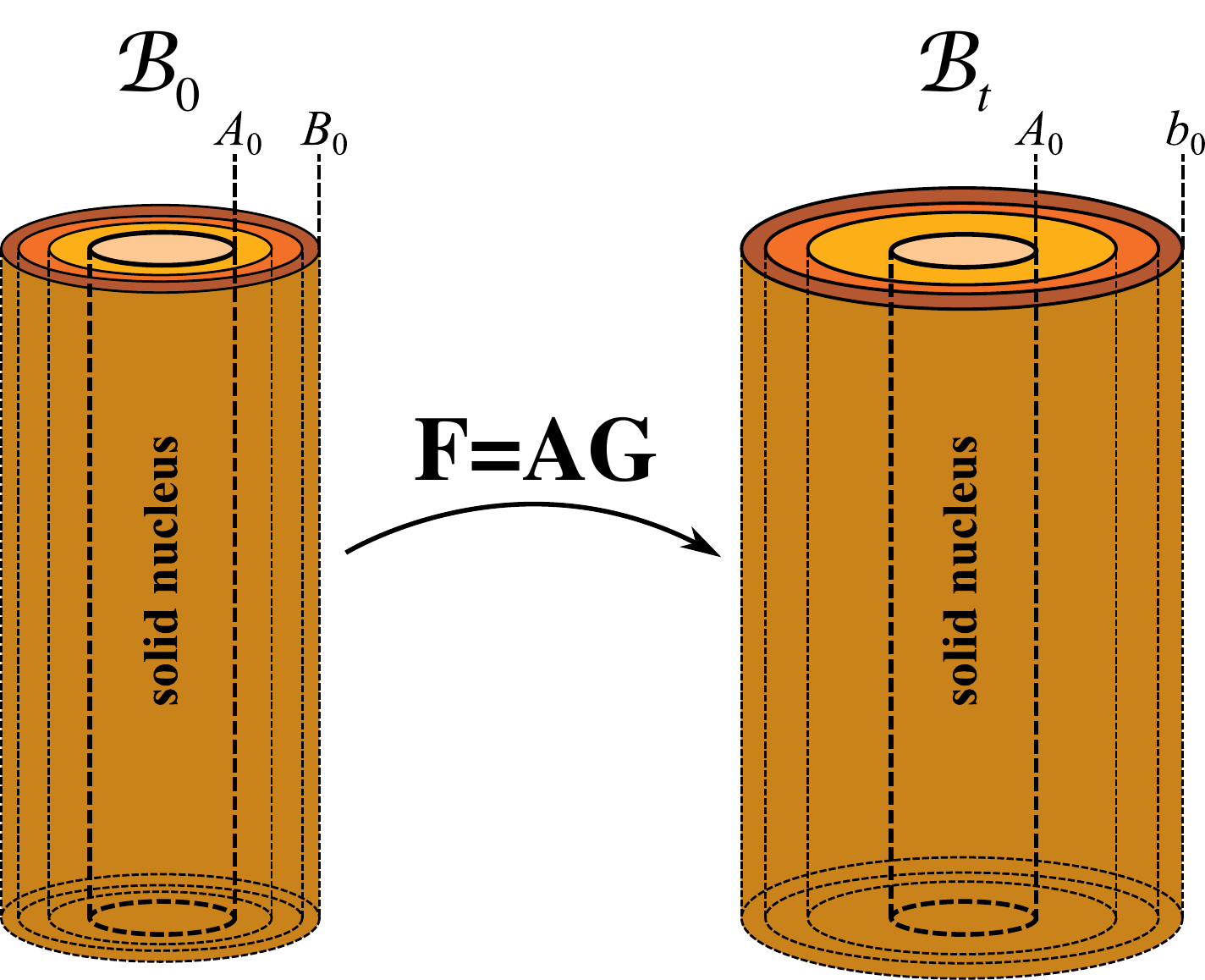}
\caption{\label{fig:Kinematics_general}Sketch of kinematic setup.}
\end{figure}

We consider a cylindrical tube, consisting of an incompressible isotropic hyperelastic material, the inner wall of which is attached
to a fixed solid nucleus, with the outer wall unconstrained (see Figure \ref{fig:Kinematics_general}). We restrict to growth and deformations only in the cross section, such that the cylindrical geometry is always maintained and there is no axial strain. Moreover, we assume that there are no external forces, so that any deformation is caused purely by growth and the elastic response. 

\begin{figure}[htpb]
\centering\includegraphics{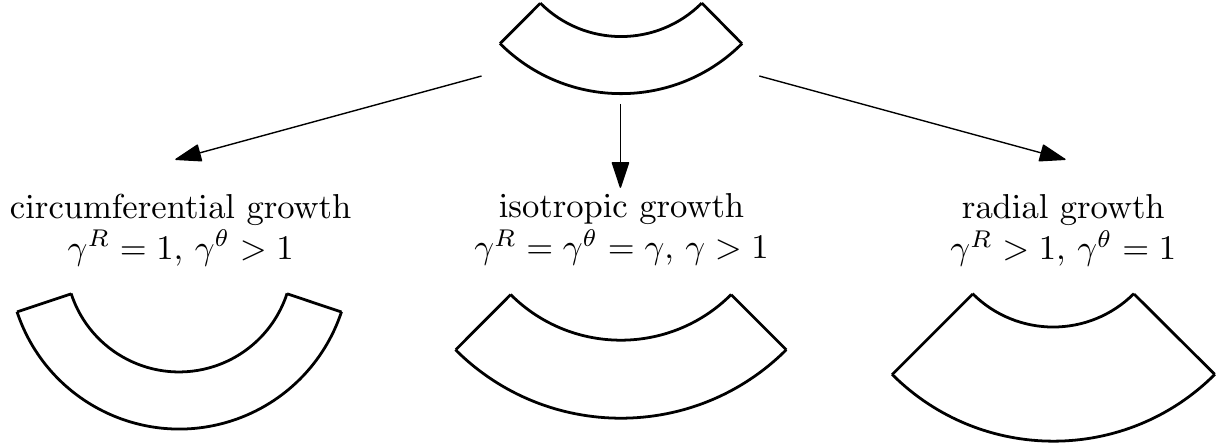}
\caption{\label{fig:growth_types}Illustration of isotropic and anisotropic
growth. }
\end{figure}

Geometrically, we work in a planar polar coordinate basis $\left\{ \mathbf{e}^{R},\mathbf{e}^{\theta}\right\} $ (the same basis vectors apply to both initial and current configurations), in which the deformation can be described by the map $\mathbf{x}:\mathcal{B}_0\to\mathcal{B}_t$ given by:
\begin{equation}
\mathbf{x}=r\left(R^{0}\right)\mathbf{e}^{R}\:.\label{eq:deformation_map_disk}
\end{equation}
For this map, the deformation gradient is
\begin{equation}
\mathbf{F}=r'\left(R^{0}\right)\mathbf{e}^{R}\otimes\mathbf{e}^{R}+\frac{r}{R^{0}}\mathbf{e}^{\theta}\otimes\mathbf{e}^{\theta}.
\end{equation}
The elastic deformation gradient takes the form
\begin{equation}
\mathbf{A}=\alpha^{R}\mathbf{e}^{R}\otimes\mathbf{e}^{R}+\alpha^{\theta}\mathbf{e}^{\theta}\otimes\mathbf{e}^{\theta}.
\end{equation}
Incompressibility requires $\det\mathbf{A}=1$; we thus define
$\alpha:=\alpha^{\theta}$, so that $\alpha^{-1}=\alpha^{r}$. We assume a diagonal growth tensor
\begin{equation}
\mathbf{G}=\gamma^{R}\mathbf{e}^{R}\otimes\mathbf{e}^{R}+\gamma^{\theta}\mathbf{e}^{\theta}\otimes\mathbf{e}^{\theta},
\end{equation}
where the difference between radial growth ($\gamma^R>1$) and circumferential growth ($\gamma^\theta>1$) is shown schematically in Figure \ref{fig:growth_types}. 
In matrix form (with the basis $\left\{ \mathbf{e}^{R},\mathbf{e}^{\theta}\right\} $ implied), we have
\begin{equation}
\mathbf{F}=\begin{pmatrix}\frac{\mathrm{d}r}{\mathrm{d}R^{0}} & 0\\
0 & \frac{r}{R^{0}}
\end{pmatrix}\:,\qquad\mathbf{A}=\begin{pmatrix}\alpha^{-1} & 0\\
0 & \alpha
\end{pmatrix}\:,\qquad\mathbf{G}=\begin{pmatrix}\gamma^{r} & 0\\
0 & \gamma^{\theta}
\end{pmatrix}\:.
\end{equation}


In the initial (stress-free) reference configuration $\mathcal{B}_{0}$, the inner
cylinder wall is located at $R^{0}=A_{0}$ and the outer wall is located
at $R^{0}=B_{0}$. From the morphoelastic decomposition $\mathbf{F}=\mathbf{AG}$,
we find $r'=\gamma^{R}/\alpha$ and $r/R^{0}=\alpha\gamma^{\theta}$.
By eliminating $\alpha$, we obtain 
\begin{equation}
r\left(R^{0}\right)r'\left(R^{0}\right)=\gamma^{R}\left(R^{0}\right)\gamma^{\theta}\left(R^{0}\right)R^{0}.\label{eq:bvp-kinematics}
\end{equation}
Imposing the boundary condition $r\left(A_{0}\right)=A_{0}$, due to 
the unmoving solid nucleus, we
can integrate \eqref{eq:bvp-kinematics} as 
\begin{equation}
r=\sqrt{A_{0}^{2}+2\int_{A_{0}}^{R^{0}}\!\!\!\gamma^{R}(\tilde{R})\gamma^{\theta}(\tilde{R})\tilde{R}\ \mathrm{d}\tilde{R}}.\label{eq:radial_map_general}
\end{equation}

\subsection{Mechanics}

Given that all deformations are diagonal in the coordinate basis considered
here, the Cauchy stress is also diagonal
\begin{equation}
\mathbf{T}=T^{RR}\mathbf{e}^{R}\otimes\mathbf{e}^{R}+T^{\theta\theta}\mathbf{e}^{\theta}\otimes\mathbf{e}^{\theta}.
\end{equation}
Let $W\left(\alpha^{R},\alpha^{\theta}\right)$ be the strain-energy density, which relates to the Cauchy stress tensor by $\mathbf{T} = \mathbf{A}W_\mathbf{A} - p\mathbf{1}$, where $p$ is the Lagrange multiplier enforcing incompressibility. In components, this reads
\begin{equation}
T^{RR}=\alpha^{r}\frac{\partial W}{\mathrm{\partial\alpha}^{r}}-p\:,\qquad T^{\theta\theta}=\alpha^{\theta}\frac{\partial W}{\mathrm{\partial\alpha^{\theta}}}-p\:/
\end{equation}
With no external loads, mechanical equilibrium requires $\text{div }\mathbf{T}=0$, which takes the form
\begin{equation}
\frac{\partial T^{RR}}{\partial r}=\frac{T^{\theta\theta}-T^{RR}}{r}.\label{eq:linear_momentum_balance}
\end{equation}
Defining $\widehat{W}\left(\alpha\right):=W\left(\alpha^{-1},\alpha\right)$, we have
\begin{equation}
T^{\theta\theta}-T^{RR}=\alpha\widehat{W}'(\alpha).
\end{equation}
In this paper we restrict analysis to a neo-Hookean strain-energy density 
\begin{equation}
\widehat{W}\left(\alpha\right)=\frac{\mu}{2}\left(\alpha^{2}+\alpha^{-2}-2\right)\:,\label{eq:neo_Hookean}
\end{equation}
for which  \eqref{eq:linear_momentum_balance} becomes 
\begin{equation}
\frac{\mathrm{d}T^{RR}}{\mathrm{d}R^{0}}  =\frac{2\mu\gamma^{R}}{R^{0}\gamma^{\theta}}\left[1-\frac{\left(R^{0}\right)^{4}\left(\gamma^{\theta}\right)^{4}}{r^{4}}\right].
\label{eq:bvp-mechanics}
\end{equation}
Along with \eqref{eq:bvp-mechanics} we impose $T^{RR}\left(B_{0}\right)=0$, i.e. the outer edge is stress-free. Equations
\eqref{eq:bvp-kinematics} and \eqref{eq:bvp-mechanics}, along with boundary condition $T^{RR}\left(B_{0}\right)=0$,  completely determine the deformation and stress state. Due to the fixed inner boundary condition, for a given growth tensor  \eqref{eq:bvp-kinematics} can be integrated separately, i.e. the deformation is determined independently from the stress, and the radial Cauchy stress is then determined by integrating \eqref{eq:radial_map_general}.
Once the radial stress component $T^{RR}$ is determined, the
circumferential component satisfies
\begin{equation}
T^{\theta\theta}=T^{RR}+\frac{2\mu r^{2}}{\left(R^{0}\right)^{2}\left(\gamma^{\theta}\right)^{2}}\left[1-\frac{\left(R^{0}\right)^{4}\left(\gamma^{\theta}\right)^{4}}{r^{4}}\right]\:.\label{eq:t2_general}
\end{equation}

Note also that for constant $\gamma^R$ and $\gamma^\theta$, these integrals may be performed analytically, giving explicit expressions for the stress and deformation in terms of the growth. As we show later, the same holds when extending from one layer to multiple layers; if the growth in each layer is constant, the stress components may be written explicitly. It is this fact that we exploit below in formulating a discretized growth dynamics. This is the main motivating reason for the fixed core geometry we consider. Under different boundary conditions, the deformation and stress would be coupled, requiring for instance a root finding exercise to determine the outer radius for which the stress boundary condition is satisfied. In such a case, the framework below applies at the expense of added computational complexity.

\subsection{\label{subsec:Growth-law-general-setup}Growth law}

We now impose a homeostasis driven growth law of the form \eqref{eq1}. In the plane polar geometry, this takes the form
\begin{equation}
\begin{aligned}\dot{\gamma}^{R} & =\left\{ K^{RR}\left[T^{RR}-\left(T^{RR}\right)^{*}\right]+K^{R\theta}\left[T^{\theta\theta}-\left(T^{\theta\theta}\right)^{*}\right]\right\} \gamma^{R}\:,\\
\dot{\gamma}^{\theta} & =\left\{ K^{\theta R}\left[T^{RR}-\left(T^{RR}\right)^{*}\right]+K^{\theta\theta}\left[T^{\theta\theta}-\left(T^{\theta\theta}\right)^{*}\right]\right\} \gamma^{\theta}\:.
\end{aligned}
\label{eq:growth_dynamics_general}
\end{equation}
Here $K^{RR}:=\mathcal{K}^{RRRR}$, $K^{R\theta}:=\mathcal{K}^{RR\theta\theta}$,
$K^{\theta R}:=\mathcal{K}^{\theta\theta RR}$, $K^{\theta\theta}:=\mathcal{K}^{\theta\theta\theta\theta}$ are the only non-vanishing components of the fourth order tensor $\mathcal{\boldsymbol{K}}$, and are assumed to be constant in space and time. 

\subsection{Discretisation approach.}
For given homeostatic stress values and components of $\mathcal{\boldsymbol{K}}$, the growth dynamics is fully defined, with the growth components evolving according to \eqref{eq:growth_dynamics_general}. Even in the simplified cylindrical geometry, this comprises a system of nonlinear partial differential equations. Moreover,  viewing the dynamics as a discrete process is still complicated by the fact that at each time step updating the growth requires knowing the stress components, which requires integration of \eqref{eq:bvp-mechanics}, which requires integration of  \eqref{eq:bvp-kinematics}, which cannot be done analytically for general spatially dependent $\gamma^R$ and $\gamma^\theta$.

However, as stated above, for constant $\gamma^R$ and $\gamma^\theta$, the integrals determining stress may be computed analytically. This suggests a discretization process whereby the annular domain is divided into discrete layers, each with constant growth, and such that the growth in each layer evolves according to averaged values of the stress. In this way, analytical expressions may be determined for both the stress and the average stress, and hence the dynamics is reduced to a set of ordinary differential equations for the growth components.
The inhomogeneity of the full model is replaced
by a piecewise homogeneous model. This preserves the key idea of inhomogeneity
(allowing, for instance, circumferential growth to be higher near
the nucleus than  away from it), but is more analytically
tractable and allows for precise statements about the long-term dynamics, stability, and qualitative investigation such as the influence of radial versus circumferential stress to the growth dynamics. 

\section{\label{sec:two_disks}Growth dynamics for 2-layer system.}

\subsection{Kinematics}

\begin{figure}[htpb]
\includegraphics[width=10cm]{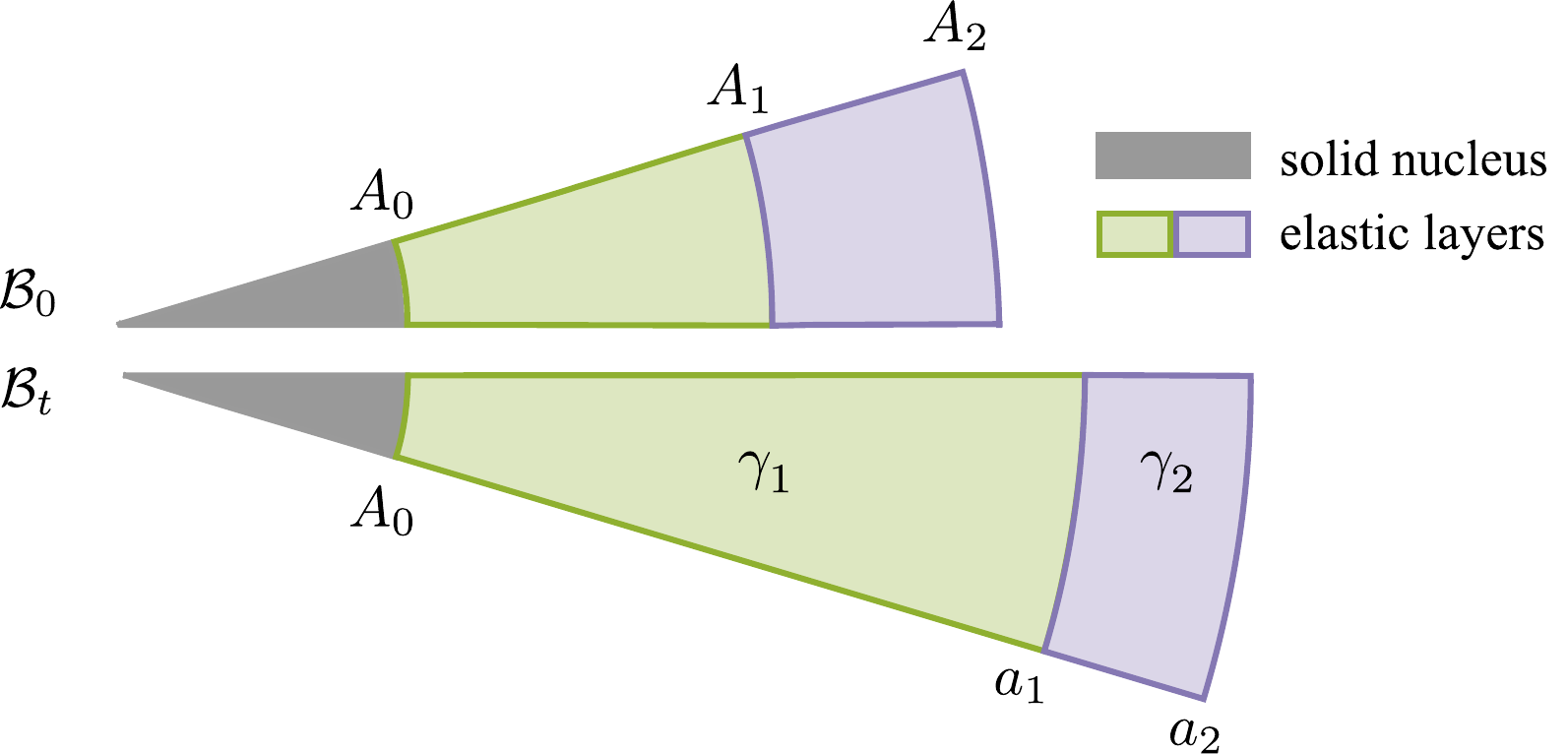}\hfill{}
\caption{\label{fig:Kinematics_two_layers}Kinematic setup for the two-layer system.
The innermost layer is attached to an unmoving nucleus ($a_{0}=A_{0}$)
and the boundary condition at the outer layer is no pressure $T^{RR}\left(A_{2}\right)=0$. }
\end{figure}

We first  consider two elastic layers attached to a solid nucleus and in perfect mechanical contact at their interface. In the initial
reference configuration $\mathcal{B}_{0}$, the inner wall has the
radial coordinate $R^{0}=A_{0}$, the middle wall at $R^{0}=A_{1}$
and the outer wall at $R^{0}=A_{2}$. In the current configuration
$\mathcal{B}_{t}$, the  same material points have coordinates are $r\left(A_{0}\right)=A_{0}$,
$r\left(A_{1}\right)=a_{1}$ and $r\left(A_{2}\right)=a_{2}$ (see Figure \ref{fig:Kinematics_two_layers}).

We impose that in the reference configuration the two annular layers
enclose the same area $\pi\Delta^{2}$ . The initial reference radii
of the two rings thus satisfy
\begin{equation}
\Delta^{2}=A_{2}^{2}-A_{1}^{2}=A_{1}^{2}-A_{0}^{2}\:.
\end{equation}

The deformation follows the same equations formulated in Section 1.1, but with piecewise homogeneous growth
\begin{equation}
\gamma\left(R^{0}\right)=\begin{cases}
\gamma_{1} & \text{if }A_{0}\leq R^{0}\leq A_{1}\\
\gamma_{2} & \text{if }A_{1}<R^{0}\leq A_{2}\:.
\end{cases}\label{eq:growth_piecewise_homogeneous}
\end{equation}
where $\gamma_1$ and $\gamma_2$ are constant. Note that our convention is to use subscript to denote different layers and superscripts for the coordinate basis index. Here, we have  imposed isotropic growth, i.e. $\gamma_{1}^{r}=\gamma_{1}^{\theta}=\gamma_{1}$
and $\gamma_{2}^{r}=\gamma_{2}^{\theta}=\gamma_{2}$. The same ideas apply for anisotropic growth, but this simplification reduces the dynamics to a 2D phase space for $\gamma_1$, $\gamma_2$. In principle, one could also have piecewise material properties and piecewise $\boldsymbol{K}$ values; however our objective is to consider the dynamics in a reduced parameter space, hence the only distinction between the layers is the different growth rates.

The deformation in each layer comes from integrating \eqref{eq:radial_map_general}, subject to $r\left(A_{0}\right)=A_{0}$ and $r\left(A_{1}\right)=a_{1}$. We obtain 
\begin{equation}
r\left(R^{0}\right)=\begin{cases}
r_{1}\left(R^{0}\right):=\sqrt{A_{0}^{2}+\gamma_{1}^{2}\left[\left(R^{0}\right)^{2}-A_{0}^{2}\right]} & \text{if }A_{0}\leq R^{0}\leq A_{1}\:,\\
r_{2}\left(R^{0}\right):=\sqrt{A_{0}^{2}+\gamma_{1}^{2}\Delta^{2}+\gamma_{2}^{2}\left[\left(R^{0}\right)^{2}-A_{1}^{2}\right]} & \text{if }A_{1}< R^{0}\leq A_{2}\:.
\end{cases}\label{eq:r_=00007Bp=00007Diecewise}
\end{equation}
Note that at $R^{0}=A_{1}$, $r$ is continuous but not differentiable.

\subsection{Mechanics}

The stress balance \eqref{eq:bvp-mechanics} determines the radial stress as
\begin{eqnarray}
&&T^{RR}\left(R^{0}\right)=\\
\nonumber &&\begin{cases}
T_{1}^{RR}\left(R^{0}\right):=T_{1}^{RR}\left(A_{1}\right)+\mu\int_{A_{1}}^{R^{0}}\frac{2}{\tilde{R}}\left(1-\frac{\tilde{R}^{4}\gamma_{1}^{4}}{r_{1}^{4}}\right)\mathrm{d}\tilde{R}, & R^{0}\in[A_{0},A_{1}]\\
T_{2}^{RR}\left(R^{0}\right):=\underbrace{T_{2}^{RR}\left(A_{2}\right)}_{0}+\mu\int_{A_{2}}^{R^{0}}\frac{2}{\tilde{R}}\left(1-\frac{\tilde{R}^{4}\gamma_{2}^{4}}{r_{2}^{4}}\right)\mathrm{d}\tilde{R}, & R^{0}\in[A_{1},A_{2}].
\end{cases}\label{eq:annulus_t1_piecewise11}
\end{eqnarray}
From \eqref{eq:t2_general}, we then obtain the circumferential stress $T^{\theta\theta}\left(R^{0}\right)$:
\begin{eqnarray}
&& T^{\theta\theta}\left(R^{0}\right)=\\
\nonumber&&\begin{cases}
T_{1}^{\theta\theta}\left(R^{0}\right):=T_{1}^{RR}\left(R^{0}\right)+\mu\frac{2r_{1}^{2}}{\gamma_{1}^{2}\left(R^{0}\right)^{2}}\left[1-\frac{\left(R^{0}\right)^{4}\gamma_{1}^{4}}{\gamma_{1}^{4}}\right], & R^{0}\in[A_{0},A_{1}]\\
T_{2}^{\theta\theta}\left(R^{0}\right):=T_{2}^{RR}\left(R^{0}\right)+\mu\frac{2r_{2}^{2}}{\gamma_{2}^{2}\left(R^{0}\right)^{2}}\left[1-\frac{\left(R^{0}\right)^{4}\gamma_{2}^{4}}{r_{2}^{4}}\right], & R^{0}\in[A_{1},A_{2}].
\end{cases}\label{eq:annulus_t2_piecewise22}
\end{eqnarray}
\begin{figure}[htpb]
\centering\includegraphics[width=1\textwidth]{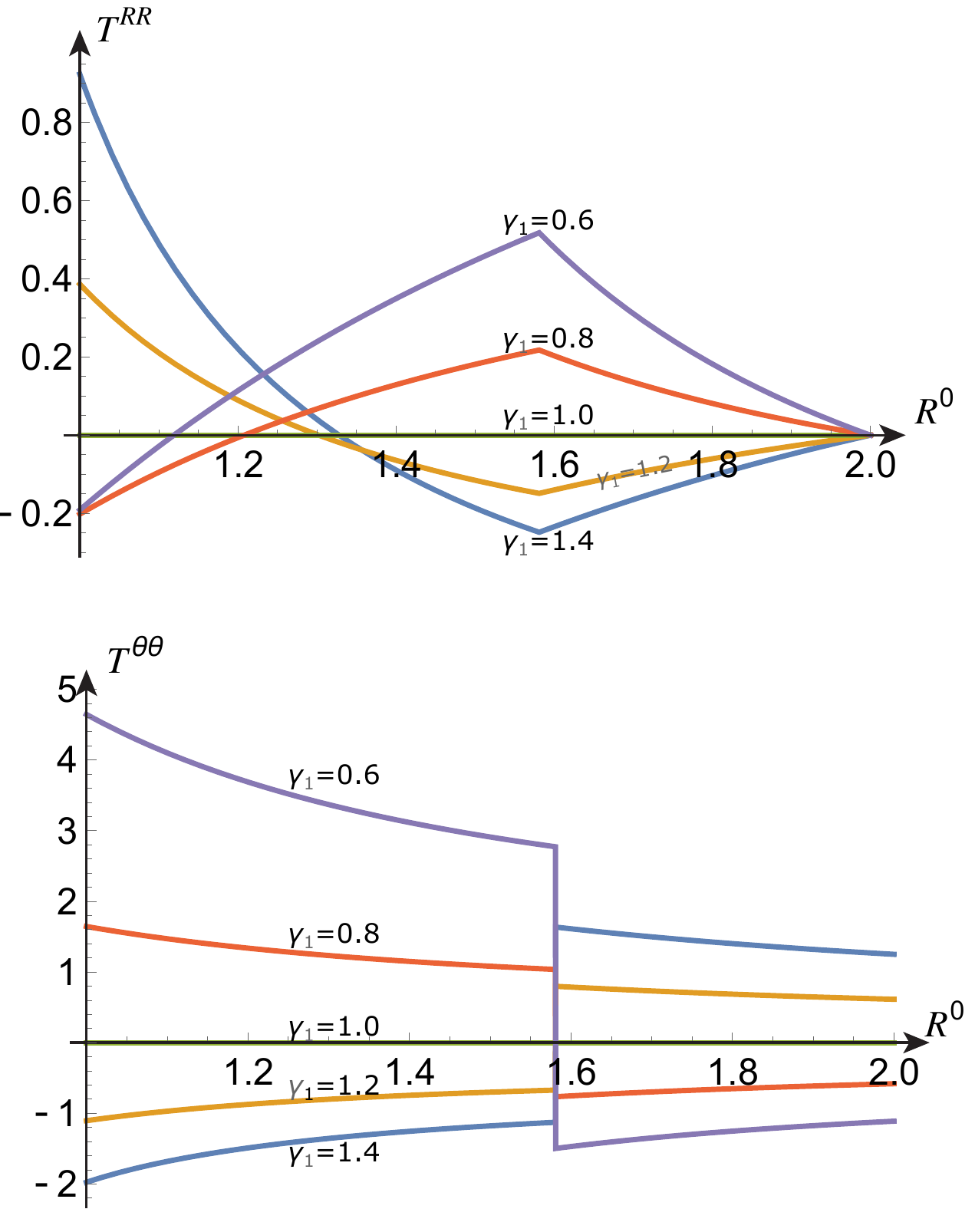}\hfill{}
\caption{\label{fig:stress_profiles-constant-gamma}Radial (top) and circumferential
(bottom) components of Cauchy stress for $A_{0}=1$, $A_{1}=\sqrt{5/2}$,
$A_{2}=2$, $\Delta=\sqrt{5/2}$, $\mu=1$, $\gamma_{2}=1$ and $\gamma_{1}$
as indicated. }
\end{figure}
The expressions $T_{1}^{RR}$ and $T_{2}^{RR}$ as well as $T_{1}^{\theta\theta}$
and $T_{2}^{\theta\theta}$ can be determined analytically as functions
of $A_{0}$, $A_{1}$, $A_{2}$, $\mu$, $\gamma_{1}$ and $\gamma_{2}$, though the exact expressions are long and have been suppressed here.

Sample stress profiles for varying values of $\gamma_1$ (with $\gamma_2=1$) are given in Figure \ref{fig:stress_profiles-constant-gamma}. With $\gamma_1>1$, the inner layer grows uniformly, hence its reference state is a uniformly expanded annulus; however, it is constrained by attachment to the core and to the ungrowing outer layer. Thus the inside of the inner layer is in radial tension (the inner edge is ``stretched'' radially to match the core), the outside is in radial compression, and the entire layer is in compression in the hoop direction. The outer layer, on the other hand, is forced to expand circumferentially to accommodate the growing inner layer and is in circumferential compression; this is balanced by a compression in the radial direction. The inverse effect occurs with $\gamma_1<1$.

\subsection{\label{subsec:Growth-law-2-layers-for-bif-diagram}Growth law}

We  define the average stresses $\overline{T_{1}}$ and $\overline{T_{2}}$,
for  both radial and circumferential stress components, as 
\begin{equation}
\overline{T_{1}}=\frac{2}{\Delta^{2}}\int_{A_{0}}^{A_{1}}T_{1}\left(\tilde{R}\right)\tilde{R}\mathrm{d}\tilde{R}\,,\qquad\overline{T_{2}}=\frac{2}{\Delta^{2}}\int_{A_{1}}^{A_{2}}T_{2}\left(\tilde{R}\right)\tilde{R}\mathrm{d}\tilde{R}.\label{eq:stress_average_two}
\end{equation}
Our approach is to modify the growth dynamics so that the (constant) growth in each layer evolves according to the averaged stress values. That is, we study the system 
\begin{equation}
\begin{aligned}\dot{\gamma}_{1} & =\gamma_{1}\left\{ K^{RR}\left[\overline{T_{1}^{RR}}-\left(T_{1}^{RR}\right)^{*}\right]+K^{\theta\theta}\left[\overline{T_{1}^{\theta\theta}}-\left(T_{1}^{\theta\theta}\right)^{*}\right]\right\} \,\\
\dot{\gamma}_{2} & =\gamma_{2}\left\{ K^{RR}\left[\overline{T_{2}^{RR}}-\left(T_{2}^{RR}\right)^{*}\right]+K^{\theta\theta}\left[\overline{T_{2}^{\theta\theta}}-\left(T_{2}^{\theta\theta}\right)^{*}\right]\right\}.
\end{aligned}
\label{eq:annulus_growth_law}
\end{equation}

Note that the isotropic growth enforces $K^{RR}=K^{\theta R}$ and $K^{\theta\theta}=K^{R\theta}$,
hence there are only two (rather than four) growth rate constants $K^{RR}$ and
$K^{\theta\theta}$. 
To further reduce the parameter space, we make the additional assumption that the homeostatic stress
values are equivalent in layers 1 and 2, that is
\begin{equation}
\left(T^{RR}\right)^{*}:=\left(T_{1}^{RR}\right)^{*}=\left(T_{2}^{RR}\right)^{*}\qquad\text{and}\qquad\left(T^{\theta\theta}\right)^{*}:=\left(T_{1}^{\theta\theta}\right)^{*}=\left(T_{2}^{\theta\theta}\right)^{*}\,.\label{eq:homeostatic-stress-equal-in-both-layers}
\end{equation}
We emphasize that while $\overline{T_{i}^{RR}}$ and $\overline{T_{i}^{\theta\theta}}$
for $i=1,2$ are averages over actual stresses according to \eqref{eq:stress_average_two},
the homeostatic values $\left(T_{i}^{RR}\right)^{*}$ and $\left(T_{i}^{\theta\theta}\right)^{*}$
for $i=1,2$ are prescribed values that may, but need not, correspond to averages of physically realizable stresses.

To facilitate the analysis, we rescale all stress quantities by a characteristic value $\sigma$, e.g. $\hat{T}^{RR}=T^{RR}/\sigma$, and rescale time as $\hat{t}=t\sigma K^{\theta\theta}$. We also introduce 
\begin{equation}
\tilde{K}:=K^{RR}/K^{\theta\theta}\qquad\text{and}\qquad \hat{T}^{*}:=\tilde{K}\left(\hat{T}^{RR}\right)^{*}+\left(\hat{T}^{\theta\theta}\right)^{*}\,.\label{eq:ktilde-tstar}
\end{equation}
The parameter $\tilde{K}$ is a measure of anisotropy of the mechanical
feedback, i.e. a weighting of the contribution of radial vs. circumferential
stress to the (isotropic) growth response.
The rescaled growth law is then
\begin{equation}
\begin{aligned}\dot{\gamma}_{1} & =\gamma_{1}\left[\tilde{K}\overline{T_{1}^{RR}}+\overline{T_{1}^{\theta\theta}}-T^{*}\right]\\
\dot{\gamma}_{2} & =\gamma_{2}\left[\tilde{K}\overline{T_{2}^{RR}}+\overline{T_{2}^{\theta\theta}}-T^{*}\right].
\end{aligned}
\label{eq:dynamical_system_two_layers}
\end{equation}
Here we have re-defined the overdot as derivative with respect
to the rescaled time, and we have dropped all hats for notational convenience. Note that all stress averages depend nonlinearly
on $\gamma_{1}$ and $\gamma_{2}$, but not on the spatial coordinate
$R^{0}$, which has been integrated out.  

\subsection{Stability analysis}

To investigate the behavior of the growth dynamics, we can now apply standard techniques of dynamical systems to \eqref{eq:dynamical_system_two_layers}; i.e. we seek equilibria satisfying
$\dot{\gamma}_{1}=0$ and $\dot{\gamma}_{2}=0$ and compute their stability. Let $\left\{ \gamma_{1}^{\text{eq}},\gamma_{2}^{\text{eq}}\right\} $ denote an equilibrium state. The nonlinear nature of the dependence of $\overline{T_{1}^{RR}}$, $\overline{T_{2}^{RR}}$, $\overline{T_{1}^{\theta\theta}}$
and $\overline{T_{2}^{\theta\theta}}$ on $\gamma_{1}$, $\gamma_{2}$
makes it difficult to compute analytically the number and
location of equilibrium states as a function of the parameters $\tilde{K}$
and $T^{*}$ and we shall use numerical methods to this end. 

For a given equilibrium state, we then perform a linear stability analysis. Let $0<\varepsilon\ll1$ and expand as
\begin{equation}
\begin{aligned}\gamma_{1} & =\gamma_{1}^{\text{eq}}+\varepsilon\overline{\gamma}_{1}+\mathcal{O}\left(\varepsilon^{2}\right),\\
\gamma_{2} & =\gamma_{2}^{\text{eq}}+\varepsilon\overline{\gamma}_{2}+\mathcal{O}\left(\varepsilon^{2}\right).
\end{aligned}
\label{eq:linear-expansion-two-layers}
\end{equation}
Introducing $\boldsymbol{\gamma}=\left(\gamma_{1},\gamma_{2}\right)$
to describe the state of the system \eqref{eq:dynamical_system_two_layers},
its linearly expanded version (to order $\varepsilon$) takes the
form 
\begin{equation}
\dot{\overline{\boldsymbol{\gamma}}}=\mathbf{J}\boldsymbol{\overline{\gamma}}
\end{equation}
 where the Jacobian matrix has entries
\begin{equation}
J_{ij}=\left[\frac{\partial\dot{\gamma}_{i}}{\partial\gamma_{j}}\right]_{\boldsymbol{\gamma}=\boldsymbol{\gamma}^{\text{eq}}}.
\end{equation}
Stability is determined in the usual way by the form of eigenvalues of $\mathbf{J}$, which are the roots of the characteristic equation 
\begin{equation}
0=(J_{11}-\lambda)(J_{22}-\lambda)-J_{12}J_{21}.
\end{equation}

\subsection{Bifurcation diagram}

\begin{figure}[htpb]
\centering\includegraphics[width=1\textwidth]{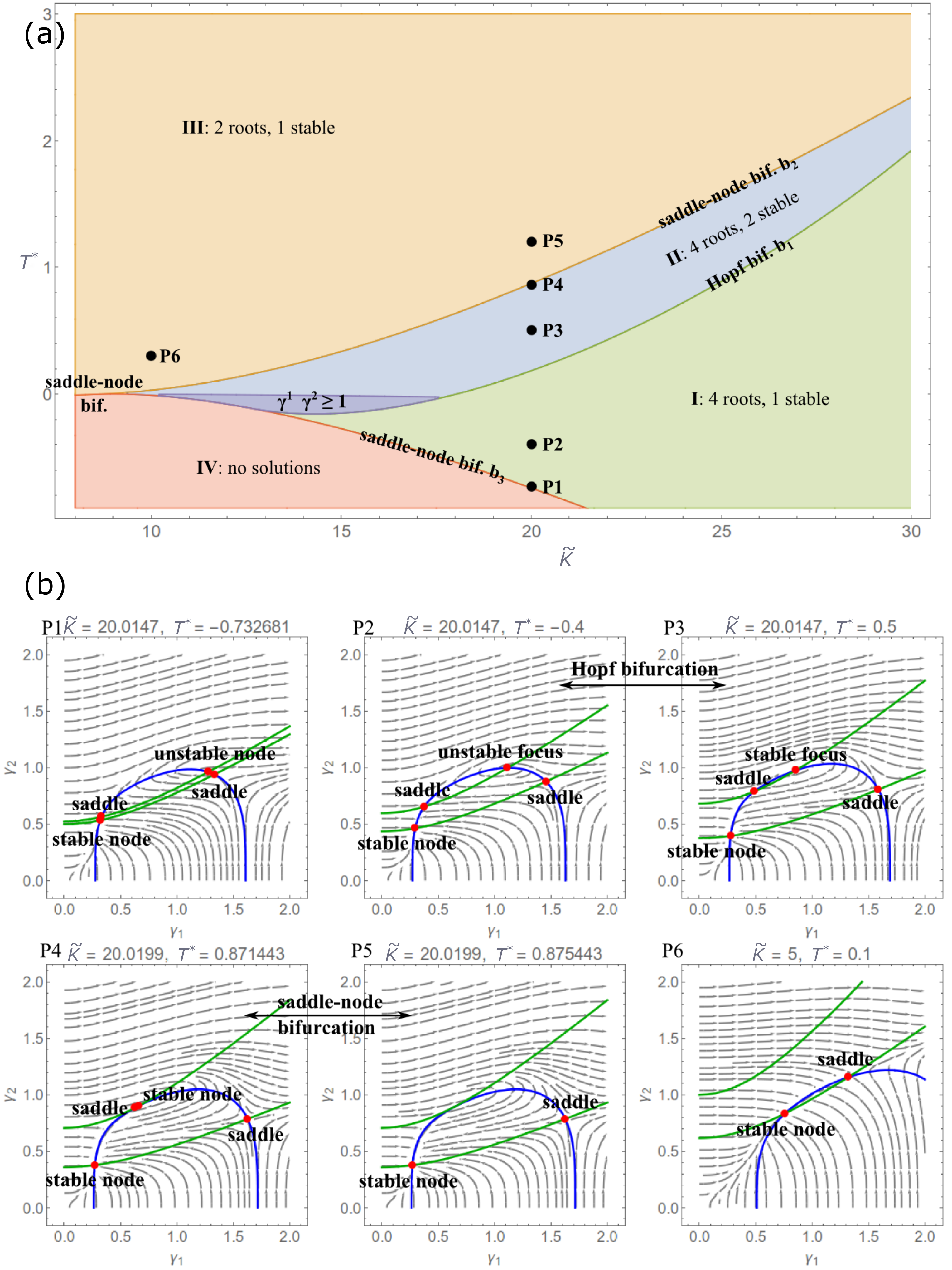}
\caption{\label{fig:two_layered_bifurcation_diagram}(a) Bifurcation diagram
for two layered actively growing piecewise homogeneous system. (b)
Equilibrium states and their dynamical characterization. Parameter values were $A_0=1$, $A_1=1.562$, $A_2=1.970$.}
\end{figure}

The number of equilibrium states and their stability depend on the values of $\tilde{K}$ and $T^*$. In Figure \ref{fig:two_layered_bifurcation_diagram}(a) we present a phase diagram that shows four regions with distinct dynamical behavior. These can be summarized as follows:
\begin{itemize}
\item \textbf{Region I} has four equilibrium states, of which one is a stable node, two are saddles, and the fourth is either an
unstable node or an unstable focus. \\
\item \textbf{Region II} has four equilibrium states: two are saddles
and the other two are either stable nodes or a stable focus and stable node. A Hopf bifurcation at the interface of Regions I
\& II transforms the unstable focus into a stable focus.\\
\item \textbf{Region III }has two equilibrium states, one of which is a
stable node, the other a saddle node. At the interface between Regions II and III, a saddle-node bifurcation occurs that annihilates the stable node and
saddle node in Region II.\\
\item \textbf{Region IV} has no equilibrium states. 
\end{itemize}
In Figure
\ref{fig:two_layered_bifurcation_diagram}(b) we show phase portraits for the selected points P1 - P5. Nullclines are plotted as blue and green curves, illustrating the appearance and disappearance of equilibrium states as categorized above. 

As is evident in Figure \ref{fig:two_layered_bifurcation_diagram}, there is a wealth of possible dynamical behavior exhibited in this system. That an idealized two-layer model with isotropic growth and equivalent homeostatic values in each layer has such a rich structure highlights a more generic complex nature of mechanically driven growth. Our intent is not to fully categorize the behavior; rather this system should be seen as a paradigm to illustrate complex dynamics. Nevertheless, several observations are in order. 

One observation from the phase portraits in Figure \ref{fig:two_layered_bifurcation_diagram}(b) is that unbounded growth is not only possible but ``common'', at least in the sense that many parameter choices and initial conditions lead to trajectories for which $\gamma_i\to\infty$. Perhaps the most natural initial condition is to set $\gamma_1=\gamma_2=1$, which corresponds to letting the system evolve from an initial state with no growth. Examining the trajectories in Figure \ref{fig:two_layered_bifurcation_diagram}(b) shows that points P1 and P2 would not evolve towards the single stable state, but rather would grow without bound.

Another point of interest is that while regions I, II and III contain stable equilibria, the stable states in Regions I and III satisfy $\gamma_{1}^{\text{eq}}\gamma_{2}^{\text{eq}}<1$. These are equilibria for which one of the layers has lost mass (at least one of the $\gamma_i<1$). Growth in both layers requires both $\gamma_i>1$, and we find that such an equilibrium only exists in a small subset of Region II, shaded dark blue in Figure \ref{fig:two_layered_bifurcation_diagram}. We further see that $T^*<0$ in the dark blue region, and $\tilde{K}$  approximately in the range 10 to 17. This implies that in order for a stable equilibrium to exist where both layers have grown, the homeostatic stress must be compressive in one or both components, and the system must respond more strongly to radial than to circumferential stress.

\paragraph{Admissible versus inadmissible homeostatic values.}
In  Figure \ref{fig:two_layered_bifurcation_diagram} we imposed the homeostatic
stress $T^*$ to be equal in each layer. Moreover, $T^*$ could take any value, and thus had no direct correspondence to a physically realizable stress state.  We now define an {\it admissible homeostatic value} as the average over a stress field that can be physically realized with the given geometry and boundary conditions. Such an admissible homeostatic stress state derives from a homeostatic growth, i.e. a given growth field $\boldsymbol{\gamma}^{*}=\left(\gamma_{1}^{*},\gamma_{2}^{*}\right)^{T}$ defines a spatially dependent stress, and averaging according to \eqref{eq:stress_average_two} then gives admissible values for the homeostatic stress:

\begin{equation}
\overline{T_{i}^{RR}}\left(\boldsymbol{\gamma}^{*}\right)\qquad\text{and}\qquad\overline{T_{i}^{\theta\theta}}\left(\boldsymbol{\gamma}^{*}\right),\qquad i=1,2\,.\label{eq:homeostatic-compatibility-2-disks}
\end{equation}
An {\it inadmissible homeostatic value} is one that cannot be expressed as an average over an actual stress, i.e. there exists
no $\boldsymbol{\gamma}^{*}$ defining $\overline{\mathbf{T}}^{*}$.

\begin{figure}[htpb]
\includegraphics[width=0.99\textwidth]{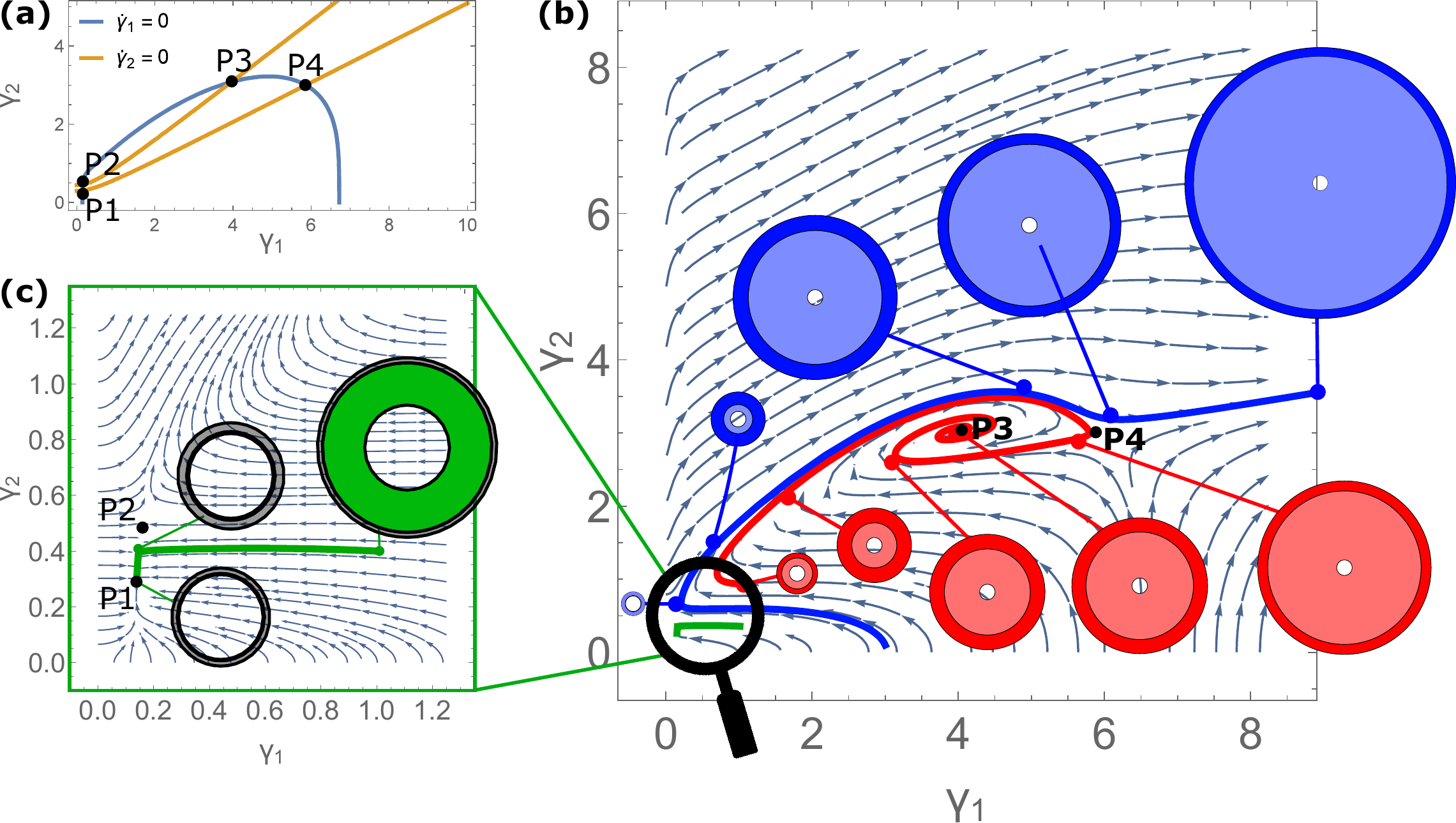}\hfill{}
\caption{\label{fig:spiral-dynamics-trajectories}Trajectories and layer sizes
for highly anisotropic growth law with admissible homeostatic state.
\textbf{(a)} Contours for $\dot{\gamma}_{1}=0$ and $\dot{\gamma}_{2}=0$
for the system \ref{fig:spiral-dynamics-trajectories}. As can be
confirmed from the stream plots \textbf{(b)} and \textbf{(c)}, there
is one stable spiral, two saddles, and one stable node. 
The saddle point P4 in (b) is the homeostatic equilibrium $\left(\gamma_{1}^{*},\gamma_{2}^{*}\right)$.
Parameters: $\mu=2$, $\Delta=\sqrt{3}$ ($A_{0}=1$, $A_{2}=\sqrt{7}$).
$\tilde{K}=23.5$. Homeostatic growth: $\gamma_{1}^{*}=5.867$, $\gamma_{2}^{*}=3$. }
\end{figure}

\paragraph{Growth law with admissible homeostatic values.}

To conclude our analysis of the two-layer system, we return to the same growth law, but for admissible homeostatic values. Due to the spatial inhomogeneity of the stress profile in the two-layer cylinder (see for instance Figure \ref{fig:stress_profiles-constant-gamma}), it is not possible to have equal homeostatic values in each layer 1 and 2. 
The growth law with admissible homeostatic values reads
\begin{equation}
\begin{aligned}\dot{\gamma}_{1} & =\gamma_{1}\left\{ \tilde{K}\left[\overline{T_{1}^{RR}}\left(\boldsymbol{\gamma}\right)-\overline{T_{1}^{RR}}\left(\boldsymbol{\gamma}^{*}\right)\right]+\left[\overline{T_{1}^{\theta\theta}}\left(\boldsymbol{\gamma}\right)-\overline{T_{1}^{\theta\theta}}\left(\boldsymbol{\gamma}^{*}\right)\right]\right\} \\
\dot{\gamma}_{2} & =\gamma_{2}\left\{ \tilde{K}\left[\overline{T_{2}^{RR}}\left(\boldsymbol{\gamma}\right)-\overline{T_{2}^{RR}}\left(\boldsymbol{\gamma}^{*}\right)\right]+\left[\overline{T_{2}^{\theta\theta}}\left(\boldsymbol{\gamma}\right)-\overline{T_{2}^{\theta\theta}}\left(\boldsymbol{\gamma}^{*}\right)\right]\right\} \,.
\end{aligned}
\label{eq:dynamics-N2}
\end{equation}
The phase space for this system is now inherently three dimensional, as the homeostatic stress values are defined by the two choices $\gamma_i^*$ as opposed to the single value $T^*$. Here we restrict our analysis to a single example, with $\gamma_{1}^{*}=5.867$, $\gamma_{2}^{*}=3$, and $\tilde{K}=23.5$, thus representing a preferred state defined by  significant growth in each layer, and with strongly anisotropic growth dynamics due to the large value of $\tilde{K}$. The dynamics are presented in Figure \ref{fig:spiral-dynamics-trajectories}. The contour plot in Figure
\ref{fig:spiral-dynamics-trajectories}(a) shows that there are in
total four equilibrium states. The streamlines and trajectory
plots in Figure \ref{fig:spiral-dynamics-trajectories}(b) and (c)
reveal that the equilibria consist of a stable spiral, two saddles, and one
stable node. It is interesting to note that P4, which is the equilibrium state at which both $\gamma_{i}^{\text{eq}}=\gamma_i^*$, is unstable; that is, the system does not remain at the equilibrium state through which the homeostatic values were defined.

Included in Figure \ref{fig:spiral-dynamics-trajectories}(b) are three sample trajectories, with the size of each layer shown at different times, and illustrative of the variety of dynamical behavior. The green trajectory quickly settles to a stable state marked by significant resorption (both $\gamma_i<1$); the blue and red trajectories sit outside the basin of attraction of P1 and show an initial period of resorption followed by significant growth. The red trajectory is in the basin of attraction of the stable focus and thus oscillates between growth and decay as it approaches the stable point at P3, while the blue trajectory, just outside the basin of attraction, ultimately grows without bound, never reaching an equilibrium state.

\section{\label{sec:N_disks}Growth of discrete $N$ layer system}

Next, we generalize the dynamical system of the previous
section from two to $N$ layers where growth and stresses are constant throughout each layer. If $N$ is sufficiently large, a system of $N$ layers can be used as a suitable spatial discretisation of a continuous growth profile on which precise statements can be obtained. In this case, we can generalize Equations \eqref{eq:dynamics-N2}  to $N$ coupled ODEs. We will analyze the stability
of this system near a homeostatic equilibrium, and show to what
extent the results obtained for $N=2$ remain unchanged as the discretisation is refined
($N$ increases), which informs the stability of the continuous ($N\rightarrow\infty$)
system. 

A major difference compared to the two-layer model
is the method to obtain homeostatic values. Previously,
homeostatic values were prescribed via the homeostatic growth values $\gamma_{1}^{*}$,
$\gamma_{2}^{*}$. In the present model, homeostatic values are obtained
by assuming the existence of  a prescribed continuous homeostatic growth profile $\gamma^{*}\left(R^{0}\right)$.
The homeostatic values $\left\{ \gamma_{i}^{*}\right\} $ are then obtained
through local averaging of the prescribed profile $\gamma^{*}\left(R^{0}\right)$
over an interval by generalizing  Equations \eqref{eq:stress_average_two}. These values are admissible by construction.
Since  growth is
taken as constant in each layer, the stresses can be determined fully
analytically and a stability analysis can then be performed. The stability
analysis will inform under which conditions the dynamical system will either
relax to a homeostatic state after a small perturbation or lead to an instability.

\subsection{Kinematics}

\begin{figure}[htpb]\centering
\includegraphics[width=8cm]{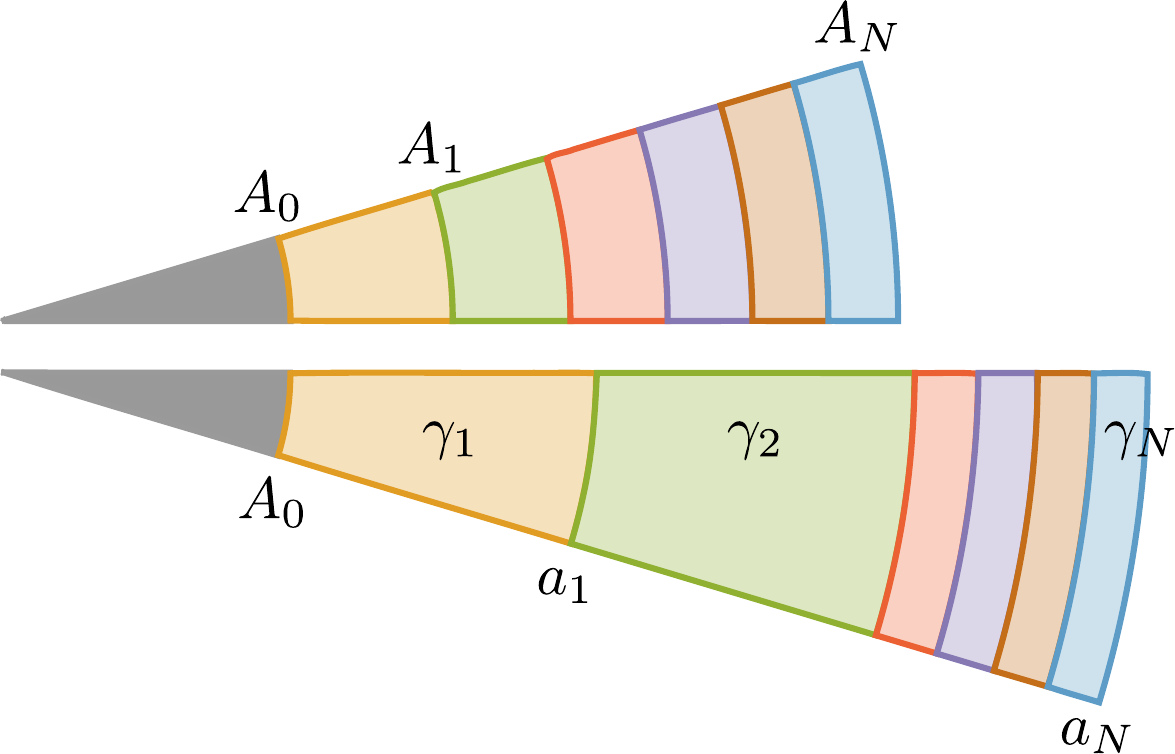}\caption{\label{fig:Kinematics_N_layers}Kinematic setup for an isotropically
growing $N$ layered system. Note that the discretization is chosen such that the areas of each layer are equal.}
\end{figure}

\begin{figure}\centering
\includegraphics[width=0.8 \textwidth]{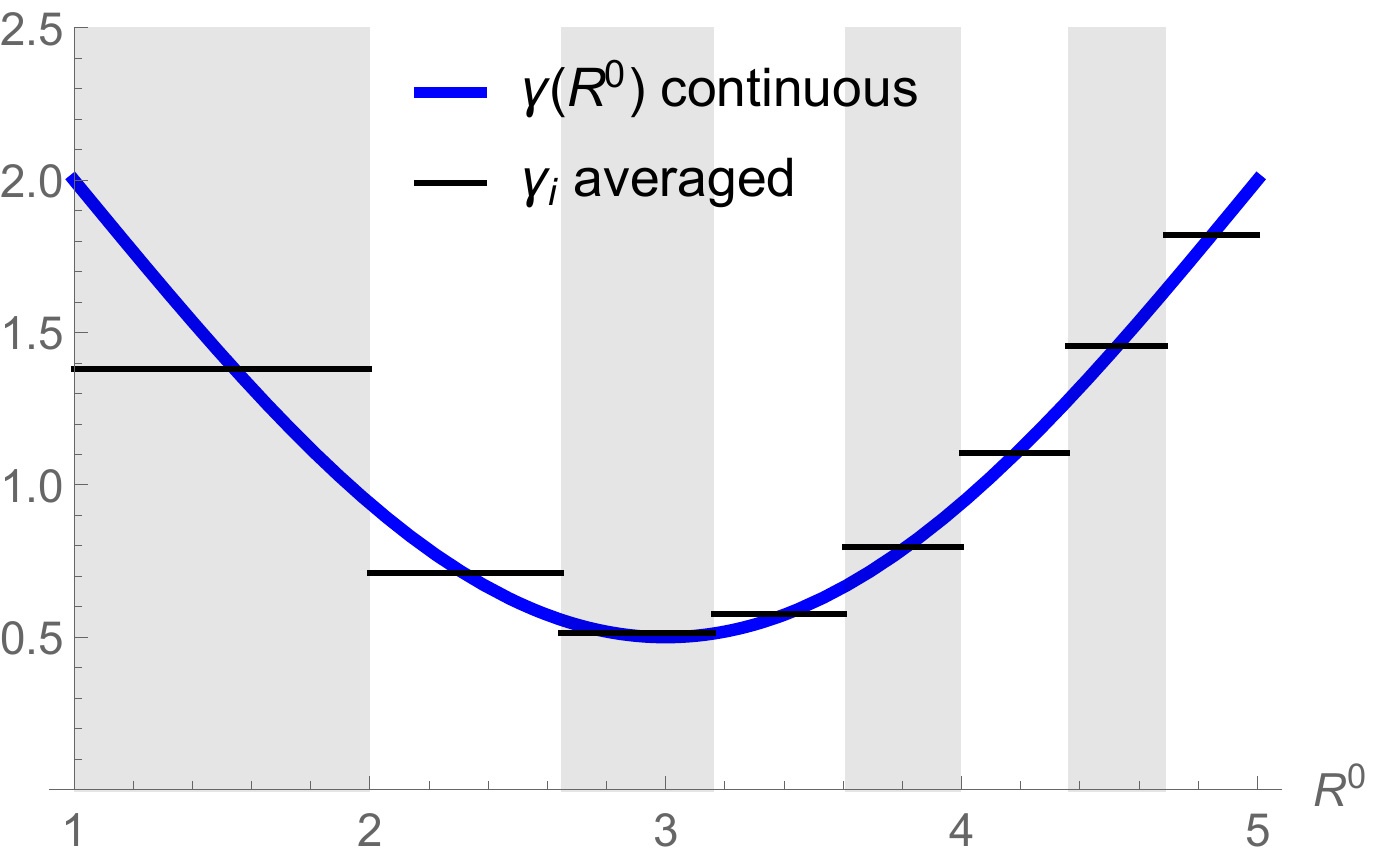}
\caption{\label{fig:averaging-gamma}Growth $\gamma$ continuous vs. averaged.
The continuous curve \eqref{eq:gamma-continuous-example} is plotted
in blue, and the average over a particular discretisation according
to \eqref{eq:gamma-piecewise} is shown by a solid piecewise constant black curve
 ($N=8$ with $A_{0}=1$, $A_{N}=5$ and $\Delta=\sqrt{3}$). }
\end{figure}

We consider $N$ perfectly connected annuli, separated by $N+1$ interfaces,
which in the initial reference configuration have the radial coordinate
values $\left\{ A_{0},A_{1},\ldots,A_{N}\right\} $ as sketched in Figure \ref{fig:Kinematics_N_layers}. 
The $K$-th annulus is defined by $A_{K-1}\leq R\leq A_{K}$ for $K\in\left\{ 1,\ldots,N\right\} $.
We choose a particular discretization so that the area between layers, $\pi\Delta^{2}$, is constant: 
\begin{equation}
A_{K}^{2}-A_{K-1}^{2}:=\Delta^{2}=\text{const.}
\end{equation}
 We can write
$A_{K}$ explicitly as 
\begin{equation}
A_{K}^{2}=A_{0}^{2}+K\Delta^{2}\,.\label{eq:Ak-explicit}
\end{equation}
Given a continuous curve $\gamma\left(R^{0}\right)$ we define the
piecewise constant growth profile by taking the average 
\begin{equation}
\gamma_{K}:=\overline{\gamma\left(R^{0}\right)}=\frac{2}{\Delta^{2}}\int_{A_{K-1}}^{A_{K}}\gamma\left(\tilde{R}\right)\tilde{R}\mathrm{d}\tilde{R},\quad K=1,\ldots,N\label{eq:gamma-piecewise}
\end{equation}
The growth value $\gamma_{K}$ is constant for all $K$.
We demonstrate the construction of the discrete profile $\left\{ \gamma_{K}\right\} $
from the continuous profile $\gamma\left(R^{0}\right)$ in Figure
\ref{fig:averaging-gamma}, in which we consider as an example the
continuous function 
\begin{equation}
\gamma\left(R^{0}\right)=2-\frac{3}{2}\sin\left(\pi\frac{R^{0}-A_{0}}{A_{N}-A_{0}}\right).\label{eq:gamma-continuous-example}
\end{equation}

Once $\left\{ \gamma_{K}\right\} $ are obtained, we compute the radial
map $r_{K}\left(R^{0}\right)$ from the discrete profile $\left\{ \gamma_{K}\right\} $.
Note that while $\gamma_{K}$ is a constant throughout the $K$-th
layer, the radial map $r_{K}$ is a function of the radial coordinate
$R^{0}$: 
\begin{equation}
r_{K}^{2}\left(R^{0}\right)=r_{K-1}^{2}\left(A_{K-1}\right)+\gamma_{K}^{2}\left[\left(R^{0}\right)^{2}-A_{K-1}^{2}\right],\qquad r_{0}^{2}\left(R^{0}\right)=A_{0}^{2}.\label{eq:radial-function-recursive}
\end{equation}
Explicitly, this implies
\begin{align}
r_{K}^{2}\left(R^{0}\right) & =A_{0}^{2}+\left(\Delta^{2}\sum_{i=1}^{K-1}\gamma_{i}^{2}\right)+\gamma_{K}^{2}\left[\left(R^{0}\right)^{2}-A_{K-1}^{2}\right].\label{eq:radial-function-explicit}
\end{align}
Notice that the recursive expression \eqref{eq:radial-function-recursive}
and the explicit expression \eqref{eq:radial-function-explicit} are
consistent with the requirement
\begin{equation}
r_{K-1}\left(A_{K-1}\right)=r_{K}\left(A_{K-1}\right),\label{eq:continuity-of-rk}
\end{equation}
which means that $r_{K}$ is continuous at the boundary layer $A_{K-1}$. 

\subsection{Mechanics}

\paragraph{Stress components.}

In the continuous version, the radial stress $T^{RR}$ is obtained from \eqref{eq:bvp-mechanics}.
The discrete version reads
\begin{equation}
\frac{\partial T_{K}^{RR}}{\partial R^{0}}=\frac{2\mu}{R^{0}}\left[1-\frac{\gamma_{K}^{4}\left(R^{0}\right)^{4}}{r_{K}^{4}\left(R^{0}\right)}\right],\qquad T_{N}^{RR}\left(A_{N}\right)=0\,.\label{eq:TRR-differential}
\end{equation}
Traction continuity  at the interfaces implies
\begin{equation}
T_{K}^{RR}\left(A_{K}\right)=T_{K+1}^{RR}\left(A_{K}\right)\,.\label{eq:continuity-of-TRR}
\end{equation}
We define $\tau^{RR}\left(R^{0}\right)$ as the indefinite integral
over the right hand side of \eqref{eq:TRR-differential} (dropping
the integration constant), 
\begin{equation}
\tau^{RR}\left(R^{0}\right):=-\mu\frac{r_{K-1}^{2}\left(A_{K-1}\right)-A_{K-1}^{2}\gamma_{K}^{2}}{r_{K}^{2}\left(R^{0}\right)}-\mu\log\left[\frac{r_{K}^{2}\left(R^{0}\right)}{\left(R^{0}\right)^{2}}\right],
\end{equation}
from which, we express the radial stress in the $K$-th layer as
\begin{equation}
T_{K}^{RR}\left(R^{0}\right)=\tau_{K}^{RR}\left(R^{0}\right)-\tau_{N}^{RR}\left(A_{N}\right)+\sum_{i=K}^{N-1}\mu\frac{A_{i}^{2}\left(\gamma_{i+1}^{2}-\gamma_{i}^{2}\right)}{r_{i}^{2}\left(A_{i}\right)}\label{eq:TRR-explicit}
\end{equation}

\begin{figure}\centering
\includegraphics[width=0.8\textwidth]{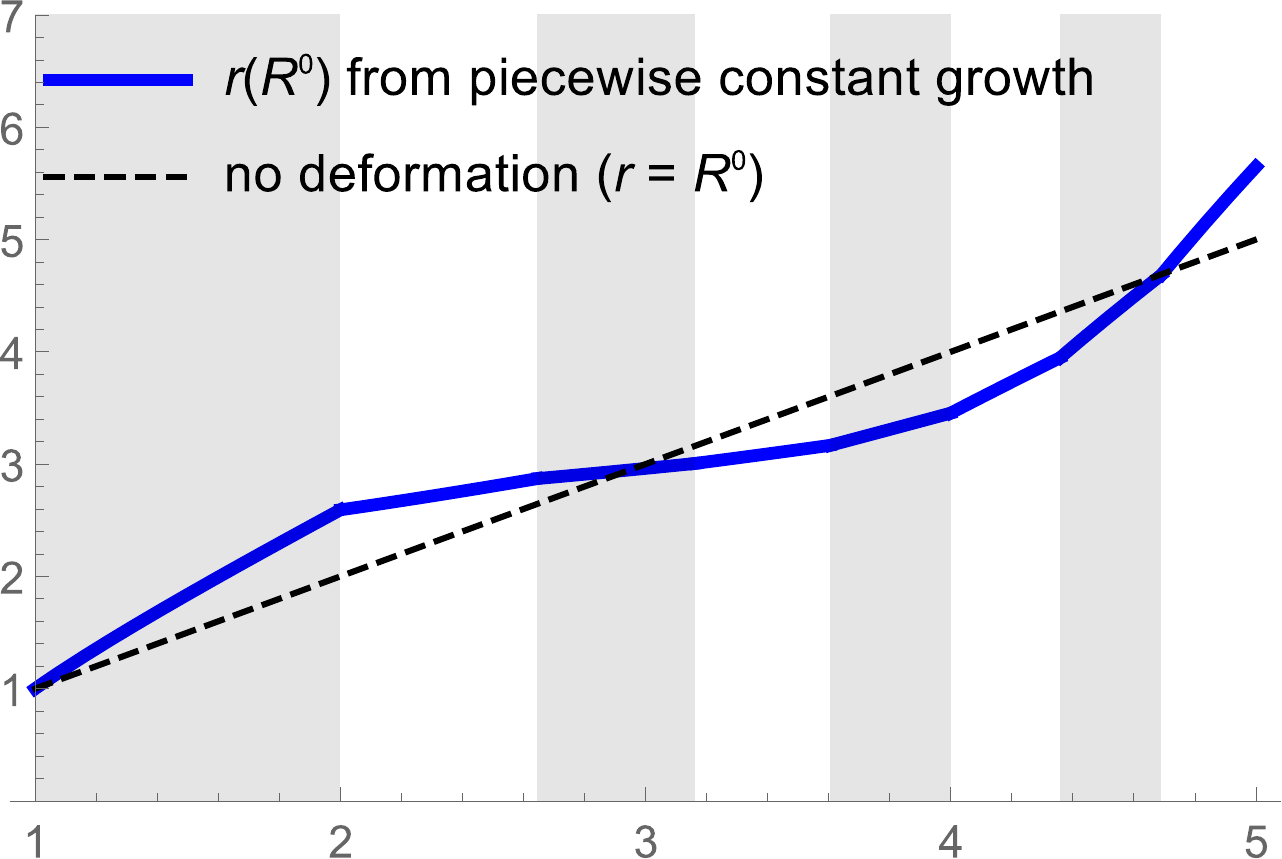}
\caption{\label{fig:radius-discrete-and-continuous}Radial function $r_{K}\left(R^{0}\right)$
for the case of discrete growth $\gamma_{i}$, computed according
to \eqref{eq:radial-function-explicit}. The dashed line represents
the case of no deformation $r=R^{0}$; everything below the dashed
line is resorption (``shrinking''), everything above this line is
growth (Parameters as in Figure \ref{fig:averaging-gamma}).}
\end{figure}

\begin{figure}[htpb]\centering
\includegraphics[width=0.8\textwidth]{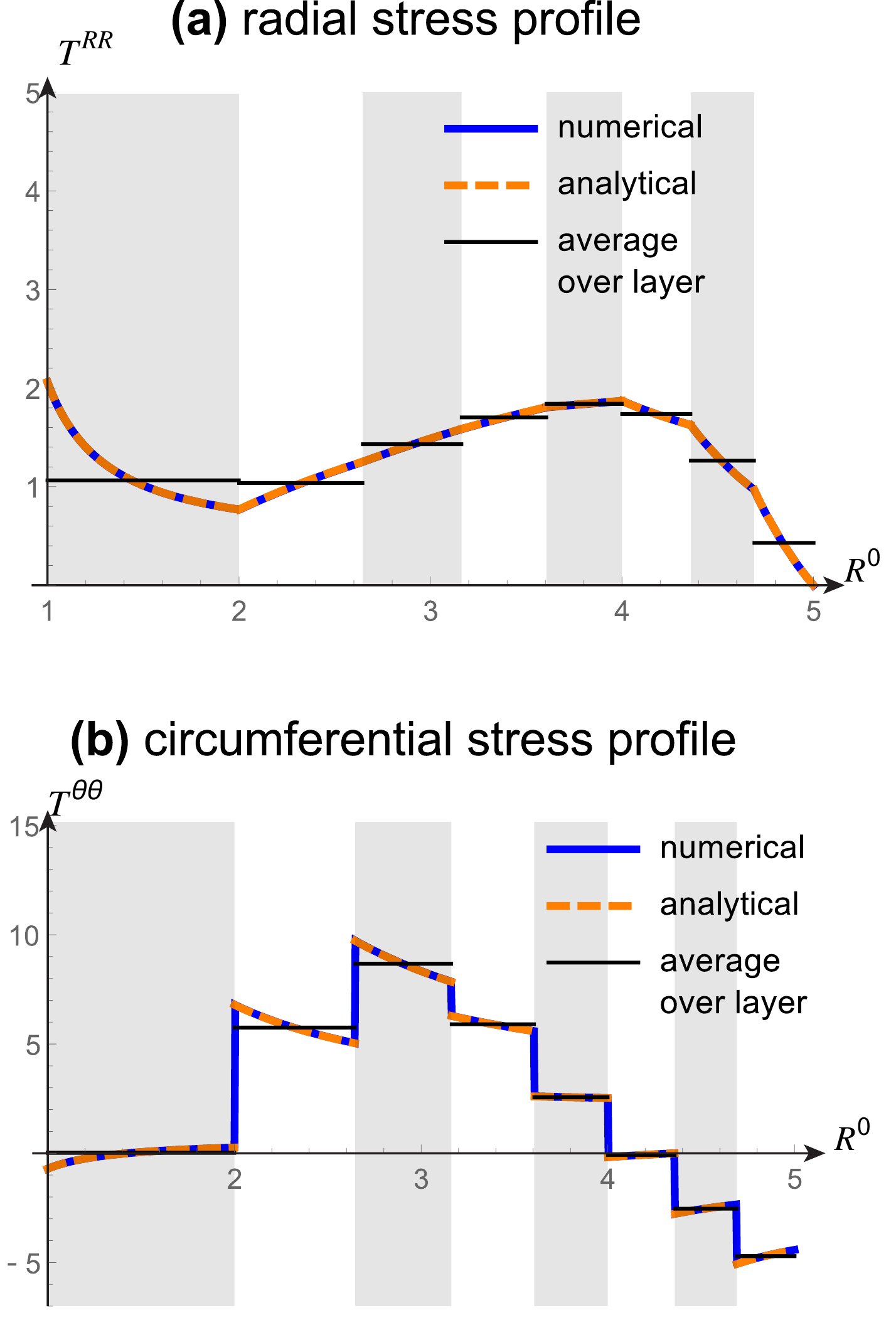}
\caption{\label{fig:discrete-stress-profile}Stress profile and stress averages
for the growth profile \eqref{eq:gamma-continuous-example}. \textbf{(a)} Radial stress profile $T^{RR}$
and average stress profile $\overline{T^{RR}}$. The analytical curve
was obtained from \eqref{eq:TRR-explicit} and the numerical curve
(for validation) was obtained from \eqref{eq:bvp-mechanics}. In both
the numerical and analytical case, the piecewise growth profile $\gamma_{i}$
according to \eqref{eq:gamma-piecewise} was used. The average stress
was computed according to \eqref{eq:TRR-average-explicit} with the
same growth profile as the other curves. \textbf{(b)} Circumferential
stress profile $T^{\theta\theta}$ and average stress profile $\overline{T^{\theta\theta}}$.
The analytical curve was obtained from \eqref{eq:TthetaTheta-discrete}
and the numerical curve (for validation) was obtained from \eqref{eq:t2_general}.
 The average stress was computed according to \ref{eq:circumferential-stress-average}.
  All other parameters are as in Figure \ref{fig:averaging-gamma}, with Young's modulus $\mu=1$. }
\end{figure}

The circumferential stress $T^{\theta\theta}$
is related to the radial stress $T^{RR}$ through \eqref{eq:t2_general}.
The discrete version of the relationship between $T^{RR}$ and $T^{\theta\theta}$
is given by
\begin{equation}
T_{K}^{\theta\theta}\left(R^{0}\right)=T_{K}^{RR}\left(R^{0}\right)+\kappa_{K}\left(R^{0}\right),\label{eq:TthetaTheta-discrete}
\end{equation}
where 
\begin{equation}
\kappa_{K}\left(R^{0}\right):=\frac{2\mu r_{K}^{2}\left(R^{0}\right)}{\gamma_{K}^{2}\left(R^{0}\right)^{2}}\left(1-\frac{\gamma_{K}^{4}\left(R^{0}\right)^{4}}{r_{K}^{4}}\right).\label{eq:kappa-discrete}
\end{equation}
Stress profiles corresponding to the growth law \eqref{eq:gamma-continuous-example} are depicted in Figure \ref{fig:discrete-stress-profile}(a) (radial) and Figure \ref{fig:discrete-stress-profile}(b) (circumferential).

\paragraph{Average stress.}

As in the two-layer case, average values for the radial and circumferential stress can be computed exactly. The average radial stress in the $K\text{-th}$ layer $\overline{T_{K}^{RR}}$
is
\begin{equation}
\overline{T_{K}^{RR}}=-\tau_{N}^{RR}\left(A_{N}\right)+\sum_{i=K}^{N-1}\mu\frac{A_{i}^{2}\left(\gamma_{i+1}^{2}-\gamma_{i}^{2}\right)}{r_{i}^{2}\left(A_{i}\right)}+\frac{2}{\Delta^{2}}\left[\nu_{K}^{rr}\left(A_{K}\right)-\nu_{K}^{rr}\left(A_{K-1}\right)\right]\label{eq:TRR-average-explicit}
\end{equation}
where $\nu_{K}\left(R^{0}\right)$ is defined as 
\begin{equation}
\nu_{K}^{rr}\left(R^{0}\right):=\mu\left[A_{K-1}^{2}-\frac{r_{K-1}^{2}\left(A_{K-1}\right)}{\gamma_{K}^{2}}\right]\log\left[r_{K}^{2}\left(R^{0}\right)\right]-\frac{1}{2}\mu\left(R^{0}\right)^{2}\log\left[\frac{r_{K}^{2}\left(R^{0}\right)}{\left(R^{0}\right)^{2}}\right].
\end{equation}

We have seen in \eqref{eq:TthetaTheta-discrete} how the circumferential
stress $T^{\theta\theta}$ relates to the radial stress $T^{RR}$.
The average over that expression is 
\begin{equation}
\overline{T_{K}^{\theta\theta}}=\overline{T_{K}^{RR}}+\overline{\kappa_{K}},\label{eq:circumferential-stress-average}
\end{equation}
We have presented an expression for $\kappa_{K}$ in \eqref{eq:kappa-discrete}.
The average over $\kappa_{K}$ is
\begin{align}
\overline{\kappa_{K}} & =\frac{2\mu\left[r_{K}^{2}\left(A_{K}\right)-\gamma_{K}^{2}A_{K}^{2}\right]}{\Delta^{2}\gamma_{K}{}^{2}}\log\left[\frac{A_{K}^{2}r_{K}^{2}\left(A_{K}\right)}{A_{K-1}^{2}r_{K-1}^{2}\left(A_{K-1}\right)}\right].\label{eq:kappa-average-explicit}
\end{align}
According to \eqref{eq:circumferential-stress-average}, the expression
for $\overline{T_{K}^{\theta\theta}}$ is the sum of $\overline{\kappa_{K}}$
(see \eqref{eq:kappa-average-explicit}) and $\overline{T_{K}^{RR}}$
(see \eqref{eq:TRR-average-explicit}). The average radial and circumferential stress components are depicted as horizontal lines in the respective
layers in Figure \ref{fig:discrete-stress-profile}(a) (radial) and Figure \ref{fig:discrete-stress-profile}(b) (circumferential). 

\subsection{\label{subsec:Generating-homeostatic-state}Generating a homeostatic state from a prescribed growth profile.}

The discretization and averaging process described above enables for a concise framework for studying growth dynamics. As the homeostatic state is defined by a growth profile -- a function that is only constrained to be positive -- a generic classification of dynamic behaviour is likely untractable. Our intent, rather, is to birefly investigate stability and the rate of convergence in terms of number of layers. For this, we restrict attention to a linear homeostatic growth profile $\gamma^{*}\left(R^{0}\right)$, characterized by  a single parameter, $C_{1}$,
\begin{equation}
\gamma^{*}\left(R^{0}\right)=1+C_{1}\left(R^{0}-A_{0}\right),\qquad C_{1}\left(A_{N}-A_{0}\right)<1.\label{eq:linear-growth-profile}
\end{equation}
Note that this growth profile satisfies $\gamma^{*}\left(A_{0}\right)=1$,
i.e. no growth at the inner boundary. 

We obtain the discrete homeostatic stress profile $\left\{ \gamma_{i}^{*}\right\} $
from the continuous profile $\gamma^{*}\left(R^{0}\right)$ by  computing
the average according to \eqref{eq:gamma-piecewise}. The homeostatic
stress  $\mathbf{T}\left(\boldsymbol{\gamma}^{*}\right)$ is
computed from the discrete homeostatic stress profile $\left\{ \gamma_{i}^{*}\right\} $
according to \eqref{eq:TRR-explicit} and \eqref{eq:TthetaTheta-discrete}.
The homeostatic values $\overline{\mathbf{T}}\left(\boldsymbol{\gamma}^{*}\right)$
are obtained as averages according to \eqref{eq:TRR-average-explicit}
and \eqref{eq:circumferential-stress-average}. It is important to
note that the homeostatic stress is generated by prescribing a growth
profile \eqref{eq:linear-growth-profile}, which by definition ensures that
the homeostatic stress is admissible. 

\subsection{\label{subsec:N-disks-stability}Growth Dynamics}

We consider a growth law that generalizes \eqref{eq:dynamics-N2} to $N$ layers. The main difference with \eqref{eq:dynamics-N2} is that  the values for homeostatic stress are obtained by the linear growth profile. 
The growth law reads
\begin{equation}
\dot{\gamma}_{K}=\gamma_{K}\left\{ \tilde{K}\left[\overline{T_{K}^{RR}}\left(\boldsymbol{\gamma}\right)-\overline{T_{K}^{RR}}\left(\boldsymbol{\gamma}^{*}\right)\right]+\overline{T_{K}^{\theta\theta}}\left(\boldsymbol{\gamma}\right)-\overline{T_{K}^{\theta\theta}}\left(\boldsymbol{\gamma}^{*}\right)\right\} ,\qquad K=1\ldots N.\label{eq:growth-dynamics-N-layers}
\end{equation}
In order to consider the stability of \eqref{eq:growth-dynamics-N-layers}
in the neighborhood of the homeostatic state, we expand growth around its equilibrium values:
\begin{equation}
\gamma_{K}=\gamma_{K}^{*}+\varepsilon\tilde{\gamma}_{K}+\mathcal{O}\left(\varepsilon^{2}\right),\qquad K=1,\ldots,N.\label{eq:gamma-near-homeostasis}
\end{equation}
To linear order in $\varepsilon$, the dynamical system simplifies
to 
\begin{equation}
\dot{\tilde{\boldsymbol{\gamma}}}=\mathbf{J}\tilde{\boldsymbol{\gamma}}.
\end{equation}
The eigenvalues of the Jacobian matrix $\mathbf{J}$ characterize the stability of
\eqref{eq:growth-dynamics-N-layers} near the homeostatic state. The
components of the $N\times N$ matrix $\mathbf{J}$ are 
\begin{align}
J_{ij} & =\left[\gamma_{i}\left(\tilde{K}\frac{\partial\overline{T_{i}^{RR}}\left(\boldsymbol{\gamma}\right)}{\partial\gamma_{j}}+\frac{\partial\overline{T_{i}^{\theta\theta}}\left(\boldsymbol{\gamma}\right)}{\partial\gamma_{j}}\right)\right]_{\boldsymbol{\gamma}=\boldsymbol{\gamma}^{*}},\quad i,j=1,\ldots,N.\label{eq:Jacobian-N-layers}
\end{align}
We characterize the stability in the neighborhood of the homeostatic
state as a function of two non-dimensional parameters: The mechanical
feedback anisotropy parameter $\tilde{K}$ and the slope of the homeostatic
growth profile $C_{1}$. The latter appears in \eqref{eq:Jacobian-N-layers}
through $\boldsymbol{\gamma}^{*}$  (see Section \ref{subsec:Generating-homeostatic-state}). 

\begin{figure}[htpb]\centering
\includegraphics[width=0.8\textwidth]{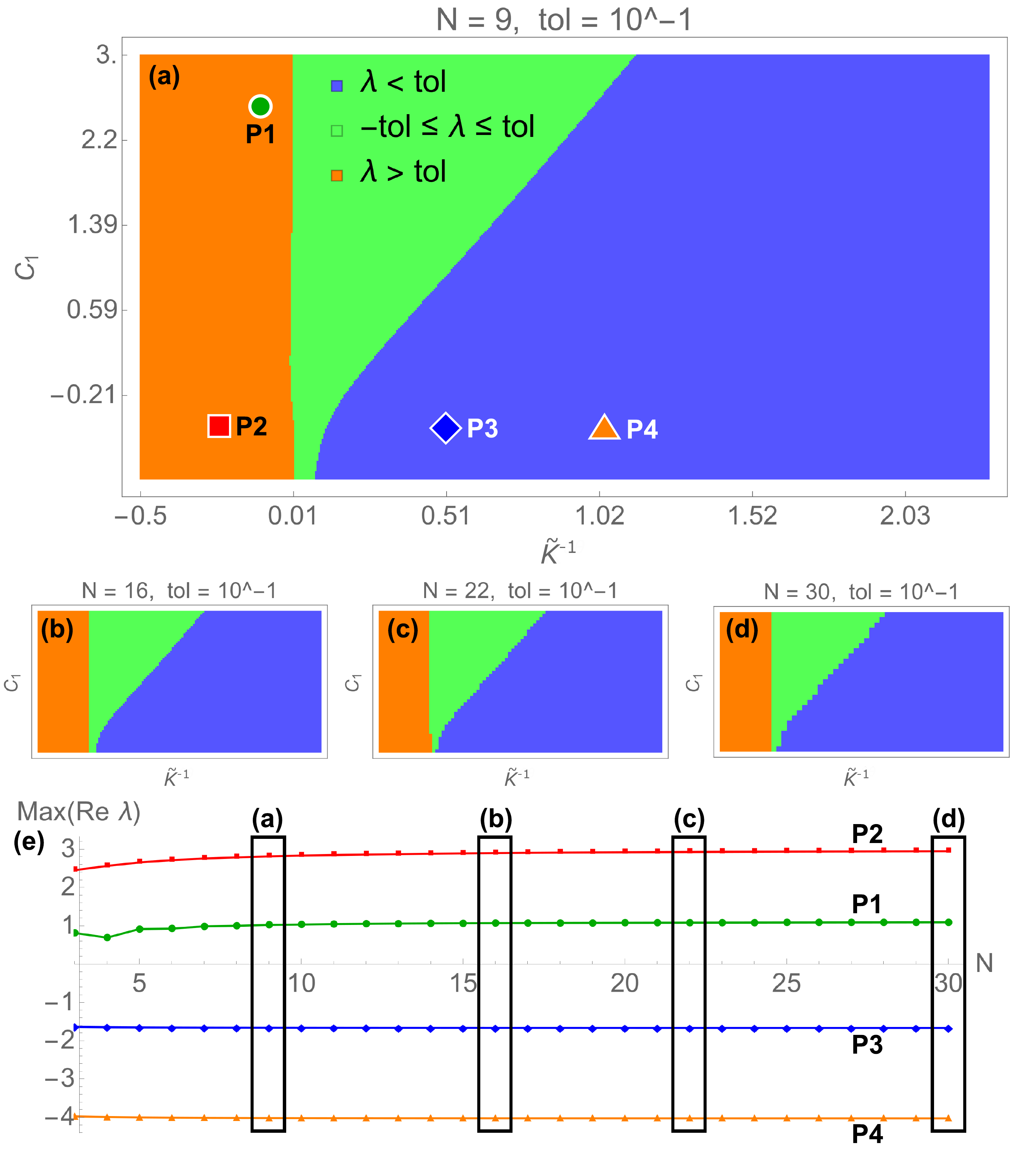}
\caption{\label{fig:Bifurcation-diagram-N-layer}Bifurcation diagram and convergence
for $N$-layered cylinder system. \textbf{(a)}\textendash}\textbf{(d)}:
The unstable (orange) and stable (blue) regions retain their shape
for increasing values of $N$. \textbf{(e)}: For a representative
sample of points P1 to P4, the convergence of the largest eigenvalue
is very good (see interpretation in text). The $\left(\tilde{K}^{-1},C_{1}\right)$
coordinates are $\text{P1}\left(0.1,2.5\right)$, $\text{P2}\left(-0.25,-0.5\right)$,
$\text{P3}\left(0.5,-0.5\right)$, $\text{P4}\left(1,-0.5\right)$.
Other parameters are $\mu_{1}=1$, $A_{0}=1$, $A_{N}=2$. 
\end{figure}

\begin{figure}[htpb]\centering\includegraphics[width=0.8\textwidth]{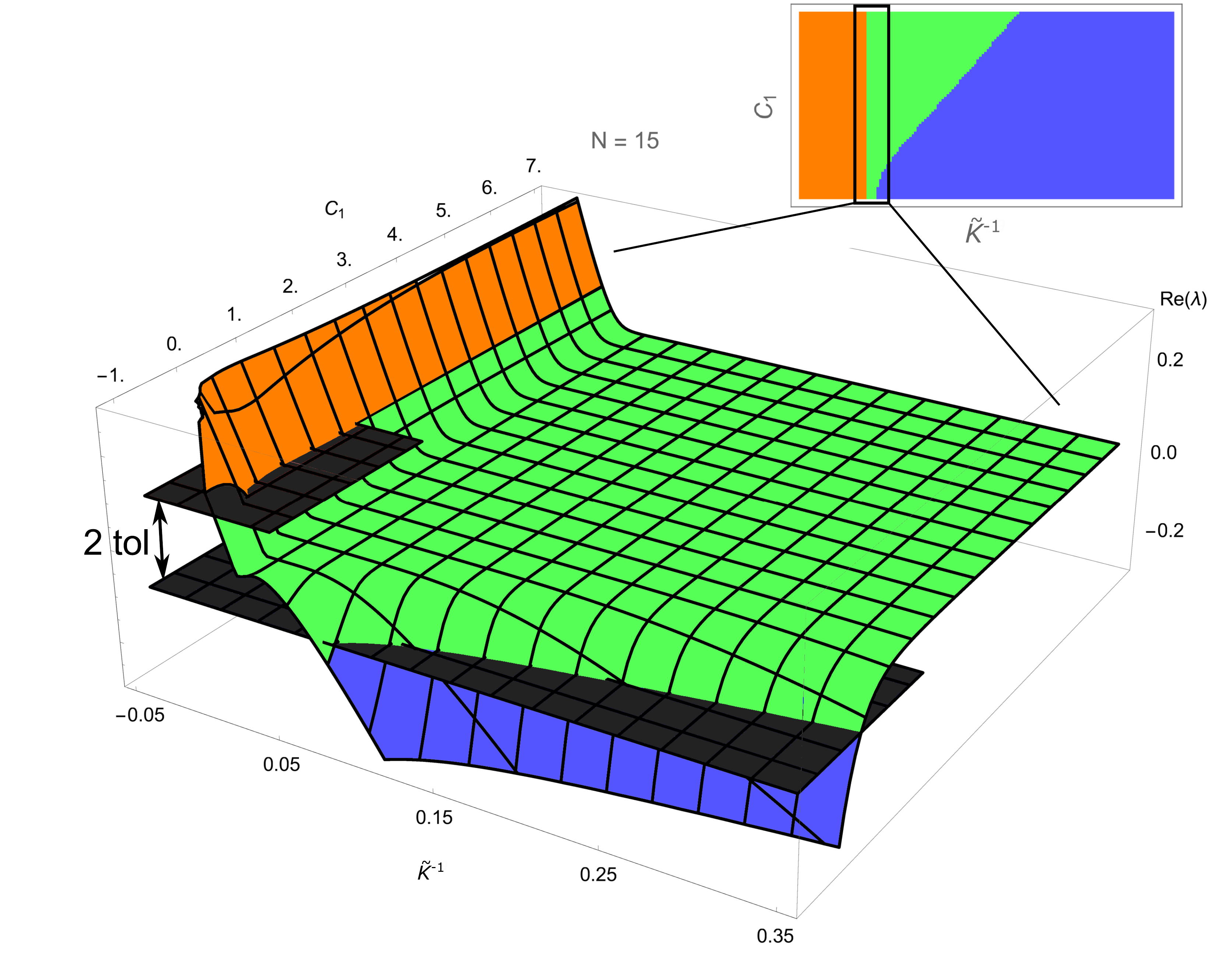}\hfill{}
\caption{\label{fig:shallow-region}Detailed depiction of the shallow (green)
region from Figure \ref{fig:Bifurcation-diagram-N-layer}. The large
shallow region has a fine structure which is an artifact of discretisation
and is not expected in the full inhomogeneous system. For this reason,
we choose a three color system in Figure \ref{fig:Bifurcation-diagram-N-layer}(a)\textendash}(d),
in which the shallow region and its fine structure are merged into
one region defined by $-\text{tol}\leq\lambda\leq\text{tol}$. The
two planes serving as upper and lower bounds of this region are depicted
in dark gray. Values above $\lambda=\text{tol}$ are stable, below
$\lambda=-\text{tol}$ are unstable. 
\end{figure}

Figure \ref{fig:Bifurcation-diagram-N-layer}(a) shows a bifurcation
diagram of the stability of the dynamical system \eqref{eq:growth-dynamics-N-layers}
as a function of $\tilde{K}^{-1}$ and $C_{1}$ for $N=9$ layers
(note that unlike in Figure \ref{fig:two_layered_bifurcation_diagram},
here we use the inverse of $\tilde{K}$ to focus on large circumferential
stress). The regions are colored
according to the largest real part of the eigenvalues $\lambda_{i}$
of $\mathbf{J}$, that is $\lambda=\text{Max(Re\ensuremath{\,\lambda_{1}}, Re\ensuremath{\,\lambda_{2}}, ... Re\ensuremath{\,\lambda_{N}})}$.
There are three parameter regions: an unstable region (orange), a
stable region (blue), and an undecidable region (green) for which
$\lambda$ is within a small tolerance of zero. This last region is included as it is typically within numerical error and its inclusion  allows to make precise statements about stability. This relatively shallow
region of $\lambda$ is further explored in Figure \ref{fig:shallow-region}
and allows us to identify the clearly stable and clearly unstable
regions of the diagram. Figure \ref{fig:Bifurcation-diagram-N-layer}(b)\textendash (e)
shows that for increasing values of $N$ (that is, a refinement of
the discretisation), the regions are practically unchanged (b\textendash d),
and that the largest eigenvalue of four selected points converges
reliably to a finite positive (P1 \& P2) or negative (P3 \& P4) eigenvalue.

The green shallow region is more explicitly visualized in Figure \ref{fig:shallow-region}.
This plot shows in the vertical axis the value of the largest real
eigenvalue computed at $\left(\tilde{K}^{-1},C_{1}\right)$ from the
Jacobian matrix \eqref{eq:Jacobian-N-layers}. The planes $\lambda=\text{tol}$
and $\lambda=-\text{tol}$ are shown in dark gray, and eigenvalues
between are assumed to be in the shallow (green) region in which stability
cannot be decided from an expansion of $\boldsymbol{\gamma}$ according
to \eqref{eq:gamma-near-homeostasis} to first order in $\varepsilon$.

Thus, we see that  there exist a region of stability, and a region of instability,
which both persist (for large enough $N$) independently of the discretisation.
A strongly anisotropic growth law ($\tilde{K}^{-1}$ close to zero
or negative) is required for the system to be unstable. We also considered the convergence as $N$ increases for a representative sample of points in the stable and unstable regions and confirmed that there was no significant change in $\lambda$. 
We expect that the stable and unstable regions represent the true behavior
of the full (inhomogeneous) system discussed in Section \ref{sec:General_disks}.
The intermediate (green) shallow region of eigenvalues has a more complicated 
structure due to the discretisation that is  not expected in the full system.

\section{Conclusion}

It is now well appreciated that growth can induce mechanical instabilities \cite{gobe05,bego05}. The related problem that we have considered in this paper is  the stability of a grown state through its slow-growth evolution.  The question is not therefore about mechanical instability but about the dynamic stability of a preferred homeostatic state. While the former is characterized by a bifurcation from a base geometry to a more complex buckled geometry, occurring on a fast elastic time-scale, the latter involves the system evolving away from a given stress state on the slow growth time-scale. In general the homeostatic state is not homogeneous, hence the issue of stability requires the analysis of partial differential equations defined on multiple configurations with free boundaries. There are no standard mathematical tools available to study this problem even for simple non-homogeneous systems. An alternative is to consider the stability of states that are piecewise homogeneous (in space). The problem is then to establish the stability of coupled ordinary differential equations describing locally homogenous states through the traditional methods of dynamical systems.  Within this framework we considered two relatively simple problems.

First, we considered the dynamical stability of a two-layer tube with different, but constant, growth tensors in each layer. We  characterized the dynamics of the full nonlinear system, and showed that the number of equilibria
and their stability varies greatly and gives rise to highly intricate
dynamics which we organized via several bifurcations. We identified 
 a parameter region where the system is stable. We found that the growth dynamics of tubular
structures in the neighborhood of the homeostatic equilibrium depends in a nontrivial way
 on the anisotropy of the growth response, and that the equilibrium
becomes unstable for highly anisotropic growth laws. This
complexity of dynamics naturally raises the question about stability of 
homeostatic equilibria for more general systems. 

Second, we showed that given a continuous law in a cylindrical geometry, we can introduce a suitable discretization of the problem that keeps all the characteristics of the continuous problem. We showed that for a linear growth law, there are
clear regions where stability and instability persist independently of
the discretisation (for sufficiently large $N$). We expect that these
regions represent the true behavior of the full inhomogeneous system.
This result allows us to characterize the stability of a morphoelastic
growing cylinder. 

While we have only scratched the surface of the complex dynamic behaviour that exists in such systems, the framework presented here provides a tool to explore growth dynamics and stability of homeostatic states and finally address some of the fundamental challenges of morphoelasticity \cite{goriely17}: What growth laws, in general, would lead to dynamically stable homeostatic states? What is the final size of a growing organism for a given growth law? What are the conditions under which growth dynamics produces oscillatory growth?

\begin{acknowledgements}
We thank Dr. Thomas Lessinnes for many useful discussions in the early stages of this project. 
 The support for  Alain Goriely by the Engineering and Physical Sciences Research Council of Great Britain under research grant EP/R020205/1 is gratefully acknowledged. 
\end{acknowledgements}

\bibliographystyle{spmpsci}      

\end{document}